\documentclass[journal]{IEEEtran}
\usepackage[utf8]{inputenc}
\usepackage[pdftex]{graphicx}
\usepackage{nicefrac}
\usepackage{balance}
\usepackage{url}
\usepackage{pgfplots}
\usepackage{multirow}
\usepackage{cite}
\usepackage{amsmath,amssymb,amsfonts,steinmetz,bm}
\usepackage[font={footnotesize}]{caption}
\usepackage{subcaption}
\usepackage{mathtools,algpseudocode,algorithm,MnSymbol}
\usepackage{enumerate}
\usepackage{textcomp}
\usepackage{xcolor,comment}
\usepackage{upgreek}
\usepackage{seqsplit}
\usepackage{balance}
\usepackage{amsthm}
\usepackage[nolist]{acronym}
\usepackage{lipsum}

\allowdisplaybreaks

\algdef{SE}[SUBALG]{Indent}{EndIndent}{}{\algorithmicend\ }%
\algtext*{Indent}
\algtext*{EndIndent}

\title{RIS-Enabled NLoS Near-Field Joint Position and Velocity Estimation under User Mobility}

\author{
Moustafa Rahal,~\IEEEmembership{Student Member,~IEEE,} 
Benoit Denis, 
Musa Furkan Keskin,~\IEEEmembership{Member,~IEEE,}
Bernard Uguen,~\IEEEmembership{Member,~IEEE,}
and Henk Wymeersch,~\IEEEmembership{Fellow,~IEEE}
\thanks{
This work was supported by the Swedish Research Council (VR grant 2022-03007), by 
the European Commission through the H2020 project RISE-6G (grant agreement no. 101017011) and Horizon Europe SNS-JU project 6G-BRICKS (grant agreement no. 101096954).\\
Moustafa Rahal and Benoit Denis are with Universit\'{e} Grenoble Alpes, CEA-Leti, F-38000 Grenoble, France (e-mails: \{moustafa.rahal,benoit.denis\}@cea.fr).}
\thanks{Musa Furkan Keskin and Henk Wymeersch are with the Department
of Electrical Engineering, Chalmers University of Technology, 41258 G\"{o}teborg, Sweden (e-mails: \{furkan,henkw\}@chalmers.se).}
\thanks{Bernard Uguen is with the Universit\'{e} Rennes 1, IETR - UMR 6164, F-35000 Rennes, France (e-mail: bernard.uguen@univ-rennes.fr).}
}

\newcommand{\vect}[1]{\boldsymbol{#1}}

\newcommand{\h}{\mathsf{H}}
\newcommand{\norm}[1]{\left\lVert#1\right\rVert}

\renewcommand{\Re}{\operatorname{Re}}
\renewcommand{\Im}{\operatorname{Im}}

\newcommand{\deltaell}{\bm{\Delta}_{\ell}}
\newcommand{\deltam}{\bm{\delta}_m}

\newcommand{\pp}{\boldsymbol{p}}

\newcommand{\cc}{\boldsymbol{c}}
\newcommand{\pz}{\pp_0}
\newcommand{\vz}{\vv_0}
\newcommand{\pd}{\pp_{\delta}}
\newcommand{\vd}{\vv_{\delta}}
\newcommand{\hh}{\boldsymbol{h}}
\newcommand{\vv}{\boldsymbol{v}}
\newcommand{\ww}{\boldsymbol{w}}
\newcommand{\uu}{\boldsymbol{u}}
\newcommand{\yy}{\boldsymbol{y}}
\newcommand{\nn}{\boldsymbol{n}}
\renewcommand{\aa}{\boldsymbol{a}}
\renewcommand{\AA}{\boldsymbol{A}}
\newcommand{\kk}{\boldsymbol{k}}
\newcommand{\kfactor}{{\mathcal{K}}}

\newcommand{\II}{\boldsymbol{I}}

\newcommand{\gammabm}{\boldsymbol{\gamma}_m}

\newcommand{\gradGam}{\nabla_{\pp} \gammabm}
\newcommand{\gradGamT}{\nabla_{\pp}^\top \gammabm}
\newcommand{\gradF}{\nabla_{\pp} \flm}

\newcommand{\conj}{ {\ast} }

\newcommand{\reposgain}{{\rm{RefPosGain}}}
\newcommand{\initposgain}{{\rm{InitPosGain}}}
\newcommand{\revel}{{\rm{RefVel}}}
\newcommand{\findposvel}{{\rm{FindPosVel}}}

\newcommand{\pphat}{\widehat{\pp}}
\newcommand{\vvhat}{\widehat{\vv}}
\newcommand{\alphahat}{\widehat{\alpha}}
\newcommand{\pdhat}{\widehat{\pp}_{\delta}}
\newcommand{\vdhat}{\widehat{\vv}_{\delta}}

\newcommand{\thetahat}{\widehat{\theta}}
\newcommand{\phihat}{\widehat{\phi}}
\newcommand{\rhohat}{\widehat{\rho}}

\newcommand{\RIS}{\text{r}}

\newcommand{\ppris}{\pp_{\RIS}}
\newcommand{\ppb}{\pp_\text{b}}
\newcommand{\ppr}{\pp_\text{r}}
\newcommand{\uuris}{\uu_\text{r}}
\newcommand{\dr}{d_\text{r}}
\newcommand{\dm}{d_{m}}
\newcommand{\qm}{q_{m}}

\newcommand{\bxil}{\bxi_{\ell}}
\newcommand{\bbl}{b_{\ell,m}}
\newcommand{\hl}{h_{\ell}}
\newcommand{\ccl}{\cc_{\ell,m}}
\newcommand{\flm}{f_{\ell,m}}

\newcommand{\etal}{\eta_{\ell}}
\newcommand{\wwl}{\ww_{\ell}}

\newcommand{\nul}{\nu_{\ell}}
\newcommand{\mubl}{\bm{\mu}_{\ell}}
\newcommand{\omegabl}{\omegab_{\ell}}
\newcommand{\Omegabl}{\Omegab_{\ell}}

\newcommand{\aat}{\widetilde{\aa}}

\newcommand{\hhtilde}{\vect{\tilde{h}}}

\newcommand{\omegab}{\bm{\omega}}
\newcommand{\Omegab}{\bm{\Omega}}

\newcommand{\bxi}{\bm{\xi}}
\newcommand{\etab}{\bm{\eta}}
\newcommand{\nub}{\bm{\nu}}
\newcommand{\Mb}{\bm{M}}
\newcommand{\bxib}{\mathbf{\Xi}}
\newcommand{\Ts}{T_s}

\newcommand{\boldzero}{{ {\boldsymbol{0}} }}
\newcommand{\boldone}{{ {\boldsymbol{1}} }}

\newcommand{\complexset}[2]{ \mathbb{C}^{#1 \times #2}  }

\newcommand{\realset}[2]{ \mathbb{R}^{#1 \times #2}  }

\newcommand{\mtcn}{\mathcal{CN}}
\newcommand{\llk}{\mathcal{L}}
\newcommand{\projrange}[1]{\boldsymbol{\Pi}_{#1}}
\newcommand{\projnull}[1]{\boldsymbol{\Pi}^{\perp}_{#1}}
\newcommand{\Imatrix}{{ \boldsymbol{\mathrm{I}} }}

\newcommand{\realp}[1]{ \Re \left\{#1\right\}}
\newcommand{\imp}[1]{ \Im \left\{#1\right\}}

\newcommand{\bW}{\boldsymbol{W}}

\newcommand{\her}{\mathsf{H}}
\newcommand{\partder}[2]{\frac{\partial{#1}}{\partial{#2}}}

\newcommand{\unit}[1]{~\mathrm{#1}}

\DeclareMathOperator*{\argmin}{arg\,min}

\DeclarePairedDelimiter\abs{\lvert}{\rvert}%

\DeclareUnicodeCharacter{0301}{\'{e}} 

\pgfplotsset{compat=1.18}

\acrodef{RIS}{reconfigurable intelligent surface}
\acrodef{BS}{base station}
\acrodef{UE}{user equipment}
\acrodef{LoS}{line-of-sight}
\acrodef{NLoS}{non-line-of-sight}
\acrodef{NF}{near-field}
\acrodef{FF}{far-field}
\acrodef{SNR}{signal-to-noise ratio}
\acrodef{SINR}{signal-to-interference-and-noise-ratio}
\acrodef{SISO}{single-input-single-output}
\acrodef{FIM}{Fisher information matrix}
\acrodef{PEB}{position error bound}
\acrodef{VEB}{velocity error bound}
\acrodef{CRLB}{Cram\'{e}r-rao lower bound}
\acrodef{RMSE}{root mean squared error} 
\acrodef{MC}{multi-carrier}
\acrodef{MIMO}{multiple inputs multiple outputs}
\acrodef{DoD}{direction of departure}
\acrodef{DoA}{direction of arrival}
\acrodef{w.r.t.}{with respect to}
\acrodef{SRE}{smart radio environment}
\acrodef{TX}{transmitter}
\acrodef{RX}{receiver}
\acrodef{QoS}{Quality of Service}
\acrodef{DL}{downlink}
\acrodef{ML}{maximum likelihood}
\acrodef{SAA}{small angle approximation}
\acrodef{SoTA}{state-of-the-art}
\acrodef{GS}{grid search}
\acrodef{CF}{closed-form}
\acrodef{EKF}{extended Kalman filter}
\acrodef{IoT}{Internet-of-Things}
\acrodef{KPI}{key performance indicator}
\acrodef{5G}{fifth-generation}
\acrodef{6G}{sixth-generation}
\acrodef{MCRLB}{Misspecified Cram\'{e}r-Rao lower bound}
\acrodef{mmWave}{millimeter wave}

\begin{document}
\maketitle

\begin{abstract}
In the context of single-\ac{BS} \ac{NLoS} single-epoch localization with the aid of a reflective \ac{RIS}, this paper introduces a novel three-step algorithm that jointly estimates the position and velocity of a mobile \ac{UE}, while compensating for the  Doppler effects observed in \ac{NF} at the \ac{RIS} elements over the short transmission duration of a sequence of \ac{DL} pilot symbols. First, a low-complexity initialization procedure is proposed, relying in part on \ac{FF} approximation and a static user assumption. Then, an alternating optimization procedure is designed to iteratively refine the velocity and position estimates, as well as the channel gain. 
The refinement routines leverage small angle approximations and the linearization of the \ac{RIS} response, accounting for both \ac{NF} and mobility effects. We evaluate the performance of the proposed algorithm through extensive simulations under diverse operating conditions with regard to \ac{SNR}, \ac{UE} mobility, uncontrolled multipath and \ac{RIS}-\ac{UE} distance. 
Our results reveal remarkable performance improvements over the \ac{SoTA} mobility-agnostic benchmark algorithm, while indicating convergence of the proposed algorithm to respective theoretical bounds on position and velocity estimation.
\end{abstract}
\acresetall
\section{Introduction}
As the \ac{5G} wireless communication systems are revolutionizing the way we connect and communicate, the industry is now gearing up for the next leap in wireless technology, namely \ac{6G}. As we set our sights on this new frontier, a new set of challenges, \acp{KPI}, and enabling technologies are emerging~\cite{6G_Shahraki2021}. The vision for \ac{6G} encompasses even higher data rates, ultra-reliable and low-latency communication, ubiquitous connectivity, and intelligent network infrastructure to support diverse applications such as autonomous systems, virtual reality, and \ac{IoT} \cite{6G_Poor, 6G_Molisch, 6G_Tao}. Meeting these ambitious goals requires breakthroughs in various enabling technologies, including advanced antenna systems, efficient spectrum utilization, edge computing, and innovative signal processing techniques \cite{6G_Bariah, 6G_Chen, 6G_Tariq}. 

Among the promising enabling technologies for the realization of \ac{6G},  \acp{RIS} have emerged as a disruptive and transformative concept~\cite{RIS_Zhao2019}. \acp{RIS}, also known as intelligent reflecting surfaces or programmable metasurfaces, are planar structures comprising numerous passive elements that can control and manipulate electromagnetic waves~\cite{RIS_Marco2020}. 
By dynamically adjusting the phase, the polarization, and even possibly the amplitude, 
of the incident signals, \acp{RIS} can actively shape the wireless propagation environment~\cite{Strinati2021}. This ability opens up a myriad of possibilities to enhance the performance and efficiency of wireless systems. \acp{RIS} can indeed contribute to optimizing signal quality, improve coverage, mitigate interference, and extend the reach of wireless networks, offering a cost-effective, energy-efficient, easy-to-deploy, and scalable 
solution compared to conventional approaches~\cite{RIS_liu2021}. The integration of \ac{RIS} into the wireless ecosystem holds great potential for changing 
the way we design, deploy, and operate future communication networks, making it a subject of significant research interest and exploration \cite{RIS_Bjornson, RIS_Wu, RIS_Basar, RIS_Alexandropoulos, RIS_Huang, RIS_Zhang}.

In addition to its impact on communication performance, the \ac{RIS} technology also holds great potential for localization applications~\cite{RIS_Henk2020}. The capability to precisely position \acp{UE} in wireless networks is indeed crucial for various emerging services, such as autonomous navigation, augmented reality, and other location-based services but also for network performance improvement~\cite{zafari2019,6G_whitepaper_Bourdoux2020}. 
For instance, by leveraging the controllable element-wise phases distribution of a reflective \ac{RIS}, 
it is possible to enhance localization continuity, accuracy, and reliability, or even simply just to make localization feasible, in restrictive scenarios and challenging environments~\cite{rahal2021_continuity}. \acp{RIS} can overcome harsh propagation conditions (e.g., \ac{NLoS}) 
or limited deployment settings for which conventional systems based on active \acp{BS} would fail~\cite{rahal2021_continuity, Kamran_Leveraging}. More generally, they can contribute to optimizing signal strength and quality to better estimate the location-dependent radio parameters required for localization~\cite{wymeersch_beyond_2020}. 
This natural synergy between \acp{RIS} and localization techniques opens up new possibilities for high-precision positioning. 
Related open research challenges are summarized in~\cite{RIS_Henk2020}. As concrete examples, reflective \acp{RIS} have been considered for parametric multipath-aided static \ac{UE} positioning in two cases: when the \ac{BS}-\ac{UE} direct path is present, namely \ac{LoS}~\cite{wymeersch_beyond_2020, Elzanaty2021, he_large_2020, zhang_towards_2020, keykhosravi2021siso}, and when it is blocked, i.e., \ac{NLoS}~\cite{rahal2021_continuity, Rahal_Localization-Optimal_RIS, liu_reconfigurable_2020}. In the former case, the \ac{BS} and \ac{UE} have at least two communication links, which facilitates the use of the traditional two-step localization approach and geometric \ac{FF} model. In contrast, when only one link is established, the previous approach might fail and a direct one-step alternative is used where we exploit the wavefront curvature in case the \ac{UE} lies in the \ac{NF} region of the \ac{RIS}. {An example of the latter is~\cite{abu2020near} where the authors used the  geometric \ac{NF} model to estimate the 3D position of the source via one \ac{RIS} acting as a lens, i.e., in reception mode.
Since operating at high frequencies and using electrically large antennas are key features of \ac{6G} systems, the Fraunhofer distance~\cite{BS2020} in such systems will extend compared to lower bands. This means that assuming a planar wave front and relying on the \ac{FF} model will result in performance degradation in most cases~\cite{CKS2023}. This model mismatch has also been studied and evaluated in terms of \ac{MCRLB}~\cite{CEG2022}. Another mismatch study has been done in~\cite{ELG2023} where authors take into consideration the implications on communication, localization and sensing all together. All this implies that if the system operates in \ac{NF}, any approximation to the \ac{FF} model would require a correction step to overcome such model mismatches.

A major drawback of the existing studies in the literature is that
only a few works so far have addressed the impact of mobility in the very context of \ac{RIS} localization, and those can be classified into two different approaches, either via snapshot estimation methods (i.e., single-epoch), where the estimated parameters are assumed to be constant over the entire transmission frame, or via tracking filters. Among the contributions that follow the former case, the authors in~\cite{keykhosravi2022siso} performed
position estimation of a mobile \ac{UE} in the \ac{FF} region of the \ac{RIS} under spatial-wideband effects, considering the presence of both the \ac{LoS} and the \ac{RIS}-induced \ac{NLoS} path.
While in~\cite{RIS_vel_est_2022}, velocity is estimated in \ac{FF} \ac{LoS} conditions, via a reflective \ac{RIS}, relying on both direct and \ac{RIS}-reflected paths.
Then, shifting away from snapshot estimation, in~\cite{Guerra2021}, a \ac{UE} transmitting a narrowband signal is tracked through filtering (i.e., including 3D velocity as estimated variable, besides 3D position), while exploiting phase and amplitude observations accounting for the curvature-of-arrival of the impinging wavefront at a \ac{RIS} in receiving mode in the \ac{NF} region.
Moreover, in~\cite{ammous2022} the authors present a tracking algorithm based on \ac{EKF} to localize \acp{UE}, in a \ac{LoS} scenario, with the aid of a reflective \ac{RIS} at the \ac{mmWave} frequency domain,
while~\cite{palmucci2023} addresses a joint \ac{RIS} reflection coefficients and \ac{BS} precoder optimization problem and estimates \ac{UE}'s trajectory in a single mobile \ac{UE} multi-\ac{RIS} MIMO scenario in the \ac{NF} regime. It is true that~\cite{Guerra2021, ammous2022, palmucci2023} take the \ac{UE} mobility into account, however, the estimation happens over multiple snapshots via a tracking filter and the mobility within individual snapshots is ignored.
In summary, the literature lacks studies where the position and/or the velocity of a mobile \ac{UE} is estimated, with a single epoch, via a reflective surface while exploiting the small-scale Doppler effects resulting from the mobility.

Therefore, in this paper, we tackle the problem of snapshot position and velocity estimation of a mobile single-antenna \ac{UE} under \ac{LoS} blockage (i.e., in \ac{NLoS}) based on multiple narrowband \ac{DL} transmissions from a single-antenna \ac{BS}, while benefiting from the collateral effects caused by the \ac{UE} mobility at a reflective \ac{RIS} in the geometric \ac{NF} propagation regime. 
%
%
More specifically, we show how the  Doppler effects induced by \ac{UE} mobility (i.e., over the time duration of the sequence of pilot signals) at the different elements of a large reflective \ac{RIS} can be taken into account and even compensated for in the estimation of position and/or velocity. 
The main paper contributions can be summarized below as follows:
\begin{itemize}
    \item \textbf{Problem Formulation for RIS-Aided 6D Near-Field Snapshot Estimation:} For the first time, we formulate the problem of single-snapshot near-field estimation of the 3D position and 3D velocity of a mobile \ac{UE} using narrowband \ac{DL} transmissions from a \ac{BS}, assisted by a \ac{RIS} in the challenging SISO scenario under \ac{LoS} blockage. 
    \item \textbf{Static \ac{UE} Position Estimation:} We propose a practical initialization procedure that provides coarse estimates for both the 3D \ac{UE} position and the complex channel gain of the \ac{RIS}-reflected path, by relying on preliminary static \ac{UE} assumption and \ac{FF} approximation, before re-injecting these results into the generic \ac{NF} \ac{RIS} response formulation for further corrections.
    \item  \textbf{Mobile \ac{UE} Position Estimation:} Leveraging the previous initialization, we also solve the \ac{UE} positioning problem with known velocity, which performs linearization and small angle approximation to compensate for position estimation residuals caused by mobility. Subsequently, we develop a global position and velocity estimation algorithm that assumes no prior information (i.e., neither on position nor on velocity) and iterates alternatively over both estimates to gradually correct their residuals.
    \item \textbf{Extensive Performance Evaluation:} We evaluate the performance of the proposed estimation framework through comprehensive simulations and benchmark against both the theoretical \acp{CRLB} (namely \ac{PEB} and \ac{VEB}) and a state-of-the-art algorithm neglecting the effect of mobility on snapshot positioning, while illustrating the impact of key parameters such as the \ac{RIS}-\ac{UE} distance in light of geometric \ac{NF} conditions, uncontrolled multipath, \ac{SNR}, \ac{UE} velocity and the prior knowledge of velocity (comparing both perfect and no velocity knowledge assumptions from the algorithm point of view).
\end{itemize}


\subsubsection*{Notations}
{Scalar variables, vectors, and matrices are respectively denoted by lower-case, lower-case bold, and upper-case bold letters (e.g., $a, \aa, \AA$) and their indices are denoted by subscripts. Moreover, $\II_N$ and $\boldzero_N$ represent the identity and the all-zero matrices of size $N\times N$, respectively. The symbols $(.)^\top$, $(.)^{*}$ and $(.)^{\mathsf{H}}$ respectively denote transpose, conjugate and hermitian conjugate, the $\mathrm{tr}(\AA)$ operator is the trace of matrix $\AA$, and $\text{diag}(\aa)$ denotes a diagonal matrix with diagonal elements defined by vector $\aa$. Furthermore, the operators $\Vert\cdot\Vert$, $\Re\{\cdot\}$ and $\Im\{\cdot\}$ represent the $l_{2}$-norm real and imaginary operators. Finally, $\mathbb{E}\{\cdot\}$ is the expectation of a random variable \ac{w.r.t.} its distribution and $\projnull{\AA} = \Imatrix - \projrange{\AA}$ the the projection into the null-space of $\AA$ with $\projrange{\AA} = \AA (\AA^\h \AA)^{-1} \AA^\h$.
}

\section{System Model}
\label{sec:SystemModel}
\begin{figure}
 \centering
 \includegraphics[width=0.99\linewidth]{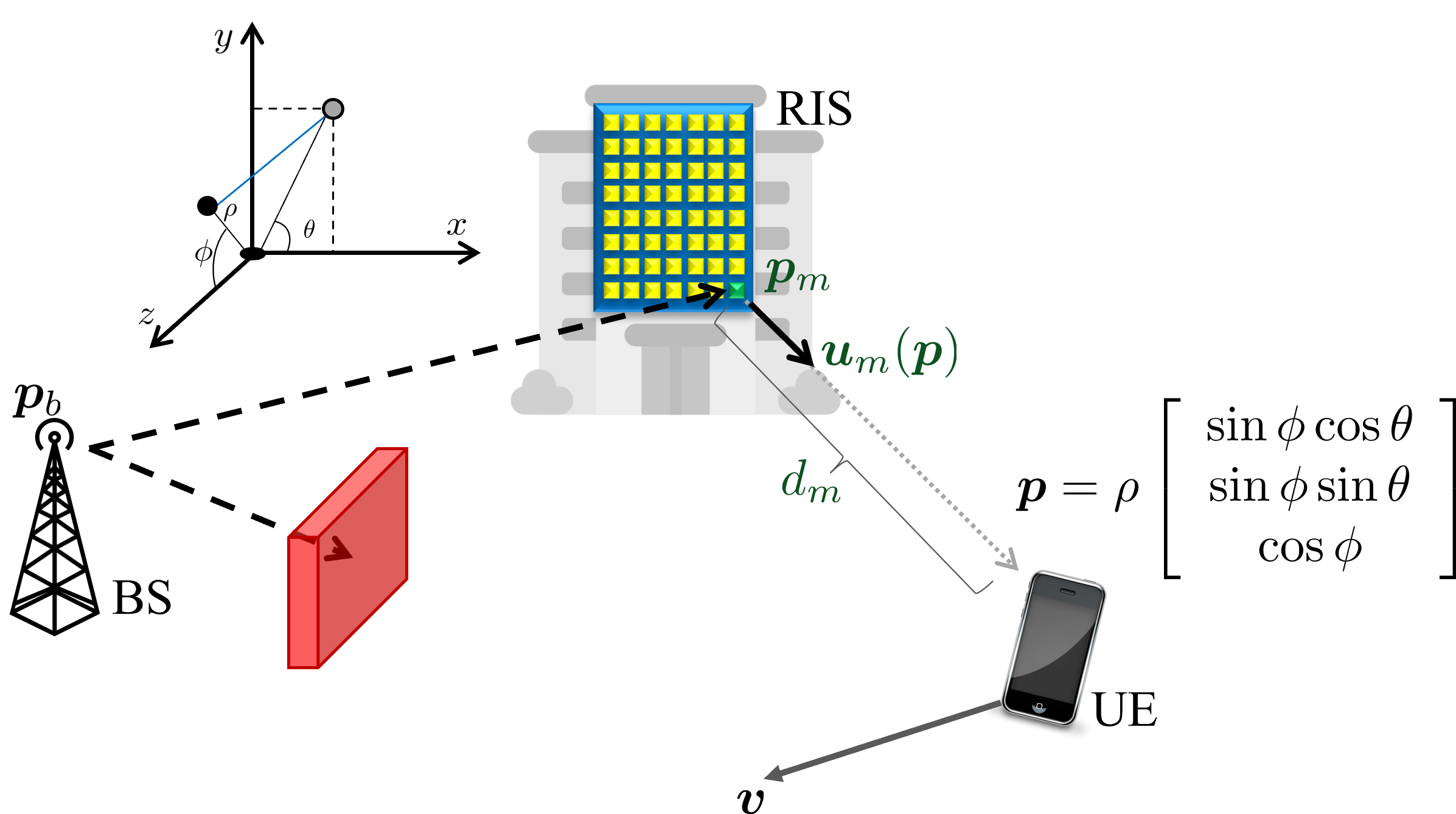}
 \caption{Considered scenario and problem geometry. The \ac{LoS} between the \ac{BS} and \ac{UE} is blocked and the \ac{UE} is in the near-field of the \ac{RIS}. 
 }
 \label{fig:model}
 \end{figure}

 We consider a SISO\footnote{Our SISO model can be generalized and extended to account for a multi-antenna arrays, i.e., MIMO (or at least MISO). In that case, the BS would illuminate the RIS using beamforming, just like in \emph{Remark 1}~\cite{pixelFailures_2024}.} \ac{DL} localization scenario with a single-antenna \ac{BS} transmitting $L$ narrowband pilot symbols, a single-antenna \ac{UE} and an $M$-element reflective \ac{RIS}, visualized in Fig.~\ref{fig:model}.
 We perform a snapshot estimation of the \ac{UE} state information, meaning that the estimation takes place at the end of each transmission frame, i.e., when all the $L$ pilots have been received. This requires the \ac{UE} state to be approximately constant during this transmission period.
The \ac{LoS} path between the \ac{BS} and the \ac{UE} is considered to be blocked, thus localization is made feasible by relying on the \ac{RIS}-reflected paths and the geometric \ac{NF} model, as in \cite{rahal2021_continuity}. In this section, we present the geometric and signal models, followed by the problem formulation. 

\subsection{Geometric Model}
The 3D positions are all expressed in the same global reference coordinates system, as follows: $\pp_{b} \in \realset{3}{1}$ is a vector containing the known \ac{BS} coordinates, $\pp_{m} \in \realset{3}{1}$ is a vector holding the known coordinates of the \ac{RIS} $m$-th element where $m \in \{1,\dots, M\}$ and $r$ is the reference element which is the \ac{RIS} center, in our case. Also, $\pp\in \realset{3}{1}$ holds the \ac{UE}'s coordinates and is expressed as $[p_x,p_y,p_z]^{\top}$ in the Cartesian coordinates system or as $[\rho, \theta, \phi]^{\top}$ in the spherical coordinates system. Fig.~\ref{fig:model} illustrates the scenario, its geometry, and the chosen coordinates axes. 
Moreover, we denote by $\vv = [v_{x}, v_{y}, v_{z}]^\top \in \realset{3}{1}$ the \ac{UE} velocity vector, and by 
$v_{m}= \vv^\top \uu_{m}(\pp)$ the radial velocities along the $m$-th \ac{RIS} element - \ac{UE} direction.
We also define $\uu_{m}(\pp) = ({\pp - \pp_{m}})/ {\dm}$, and $\dm = \norm{\pp_{m} - \pp}$.


\subsection{Signal Model}
The position of the mobile \ac{UE} at measurement instance $\ell$ (i.e., $\ell$-th pilot symbol) for $\ell \in \{1,\dots, L\}$ follows a constant velocity model, i.e. $\pp_\ell= \pp + \vv \ell \Ts$ where $\vv \ell \Ts$ denotes the displacement induced by the constant\footnote{The  constant velocity model is reasonable for sufficiently low velocities (to be detailed later) during the snapshot measurement of duration $L T_s$.} velocity vector $\vv$.
Then, leveraging the model in~\cite{abu2020near}, the received signal model at instance $\ell$ can be expressed in \ac{NF} as\footnote{We assume the absence of uncontrolled multipath in algorithm derivation \cite{keykhosravi2022siso,LOS_NLOS_NearField_2021}. In the numerical results of Section~\ref{sec:Simulations}, we will investigate the impact of uncontrolled multipath on estimation performance.}
\begin{align}
    y_\ell &=  \alpha \, \aa^{\top}(\pp_\ell) \Omegabl \aa(\ppb) s_\ell + n_\ell, \label{eq:obs_model}
\end{align}
where $\pp_\ell$ is a function of $\pp$ and $\vv$,
and $s_\ell$ denotes the transmitted pilot (taken as $s_\ell = 1, \, \forall \ell$ hereafter for ease of exposition), and $\alpha \in \mathbb{C}$ is the complex channel gain that involves the impact of the transmit power, attenuation in both \ac{BS}-\ac{RIS} and \ac{RIS}-\ac{UE} paths and global phase offset, and it is modeled as\footnote{Amplitude variations across the \ac{RIS} due to differences in distances or incident angles are ignored in this study. }~\cite{wymeersch_beyond_2020}
\begin{align}
    \alpha = \frac{\lambda^2\sqrt{PG_t G_r}}{(4\pi)^2\norm{\pp_r - \pp}\norm{\pp_r - \pp_b}} \exp(j\psi) \,,
\end{align}
where $\psi$ represents the global phase offset, $\lambda$ is the signal wavelength, $P$ denotes the transmit power, and $G_t$ and $G_r$ are the transmit and receive antenna gains, respectively.
Additionally, $\Omegabl = {\rm{diag}}(\omegabl)$ with $\omegabl \in \mathbb{C}^{M\times1}$ denoting the vector of controlled \ac{RIS} element-wise reflection coefficients, and  $n_\ell \sim \mtcn(0, N_0 W n_f)$ is an independent and identically distributed additive zero-mean normal noise with power spectral density $N_0$, bandwidth $W=1/T_s$, and noise figure $n_f$. In \eqref{eq:obs_model}, the element-wise \ac{RIS} response model can be expressed as follows:\footnote{{The published version  incorrectly considered $\norm{\ppr -  (\pp + \vv  \ell\Ts)}$ instead of $\norm{\ppr -  \pp}$ (which in turn can be absorbed into $\psi$) in the generative model. This in no way affects the methodology. This error is now fixed throughout this manuscript, though the results still rely on the incorrect generative model. }} 
\begin{align}
    [\aa(\pp_\ell)]_{m} &=  \exp\left(-j\frac{2\pi}{\lambda} \flm (\pp, \vv)
    \right)~, \label{eq:nf_steer}\\
    \flm (\pp, \vv) &= \norm{\pp_{m} - (\pp + \vv  \ell\Ts)}  - \norm{\ppr -  \pp} 
    ~,\label{eq:flm_undeveloped}\\
    & \approx \dm - \dr
    +
     \uu^\top_{m}(\pp)\vv \ell \Ts~,\label{eq:flm}
\end{align}
where~\eqref{eq:flm} is obtained from~\eqref{eq:flm_undeveloped} after a Taylor series expansion and dropping higher-order terms (See Appendix~\ref{app:flm}), and used as the generative physical model throughout the manuscript. It is worth noting that $\aa(\pp_{b})$, which represents the RIS response vector corresponding to signals arriving from the \ac{BS}, remains independent of $\ell$ since both the \ac{BS} and the RIS are stationary entities.
%
To have a compact model for the observations, we stack \eqref{eq:obs_model} over $L$ symbols and define the observation  vector $\yy \triangleq [y_1 \ \cdots \ y_L]^\top \in \complexset{L}{1}$, leading to
\begin{align}
    \yy &= \alpha \, \hh(\pp, \vv) + \nn ~,\label{eq:yy}\\
    \hh(\pp, \vv) &\triangleq [ h_1(\pp,\vv) \ \cdots \ h_L(\pp,\vv) ]^\top \in \complexset{L}{1}, \label{eq:hl_pv_vec}\\
    h_{\ell}(\pp,\vv) &\triangleq \wwl^\top \aa(\pp_\ell) \label{eq:hl_pv} ~,
\end{align}
where $\nn \triangleq [n_1 \ \cdots \ n_L]^\top$ and $\wwl \triangleq \omegabl \odot \aa(\ppb)  \in \mathbb{C}^{M\times 1}$.
\subsection{Problem Description and Maximum-Likelihood Estimator}
Given the \ac{NLoS} \ac{RIS}-induced observations \eqref{eq:yy} at the mobile \ac{UE}, our goal is to perform joint estimation of its position $\pp$ and velocity $\vv$, along with the nuisance parameter $\alpha$. 
For this problem, the \ac{ML} estimator of the unknown parameter vector $\vect{\zeta} = [\pp^\top, \vv^\top, \alpha]^{\top}$ can be expressed as
%
\begin{align}
    ( \alphahat  ~, \pphat ~, \vvhat)
    &= \arg \min_{\alpha, \pp, \vv} 
    \lVert \yy - \alpha \hh(\pp, \vv)  \rVert ^2,
    \label{eq:ml_pos_vel_gain}
\end{align}
which involves a high-dimensional search.
Alternatively, the ML estimator of $\pp$ and $\vv$ can be obtained by the \ac{CF} estimation of $\alpha$ in \eqref{eq:ml_pos_vel_gain} as
\begin{align} \label{eq:alphahat_update}
    \alphahat = \frac{  \hh^\her(\pp, \vv) \yy }{\norm{\hh(\pp, \vv)}^2} ~.
\end{align}
Plugging \eqref{eq:alphahat_update} back into \eqref{eq:ml_pos_vel_gain}, we obtain the concentrated cost function 
\begin{align}\label{eq:ml_pos_vel}
		(\pphat,\vvhat) = \arg \min_{\pp, \vv}  \norm{\projnull{\hh(\pp, \vv)} \yy }^2 ~,
\end{align}
leading to a 6D optimization problem. In the subsequent sections, our focus will be directed towards solving \eqref{eq:ml_pos_vel_gain}, and when convenient, its concentrated form \eqref{eq:ml_pos_vel}, with an emphasis on achieving low computational complexity. 
%


\section{Proposed Method}
\label{sec:proposed_methods}

In this section, we will elaborate on 
the overall procedure of estimating the 6D position and velocity parameters of the mobile \ac{UE} in the absence of prior knowledge.

\subsection{Top-Down Description}
\label{subsec:top_down}

The overall method is described in Algorithm \ref{alg:overall}. 
The methods commence by assuming a static \ac{UE} with unknown position, and then iterates between velocity and position estimations 
until the objective function~\eqref{eq:ml_pos_vel_gain} converges. 
A final, Quasi-Newton, refinement routine is performed to better estimate the positional, velocity, and gain parameters.
The subroutines $\initposgain(\yy)$ and $\revel(\yy, \vvhat, \pphat, \alphahat)$ will be described in Section \ref{subsec:PosEstimate}, and the subroutine  $\reposgain(\yy, \vvhat, \pphat, \alphahat)$ will be detailed in Section \ref{subsec:VelEstimate}.
\begin{algorithm}
	\caption{6D Estimation in RIS-Aided Near-Field Localization: $(\alphahat, \pphat, \vvhat) = \findposvel(\yy)$}
	\begin{algorithmic}[1]
	    \State \textbf{Input:} Observation $\yy$ in \eqref{eq:yy}.
	    \State \textbf{Output:} Position, velocity and gain estimates $\alphahat, \pphat, \vvhat$.
	    \State Initialize the velocity estimate as $\vvhat = \boldzero \unit{m/s}$.
	    \State Initialize the position and gain estimates 
	    \begin{align}
	        (\pphat, \alphahat) = \initposgain(\yy) ~. \nonumber
	    \end{align}
	    \State \textbf{While} the objective in \eqref{eq:ml_pos_vel_gain} does not converge
	     \Indent
	        \State Update the velocity and gain estimates 
	        \begin{align}
	            (\vvhat,\alphahat) \leftarrow \revel(\yy, \vvhat, \pphat, \alphahat) ~. \nonumber
	        \end{align}
	        \State Update the position and gain estimates 
	        \begin{align}
	            (\pphat, \alphahat) \leftarrow \reposgain(\yy, \vvhat, \pphat, \alphahat)~. \nonumber
	        \end{align}
	     \EndIndent
	     \State \textbf{end while} 
	     \State Perform 6D gradient descent search, starting from $(\pphat, \vvhat)$.
	     \State Update $\alphahat$ via \eqref{eq:alphahat_update}.
	\end{algorithmic}
	\normalsize
    \label{alg:overall}
\end{algorithm}

\subsection{Position Estimate with Known Velocity}
\label{subsec:PosEstimate}

The proposed estimation algorithm  is an extension to that proposed in~\cite{abu2020near}. 
%
Injecting the estimate of $\alpha$ from  
\eqref{eq:alphahat_update} back into~\eqref{eq:ml_pos_vel_gain}, the ML estimate of $\pp$ can be computed as 
\begin{align}\label{eq:ml_pos}
		\pphat = \arg \min_{\pp}  \norm{\projnull{\hh(\pp, \vv)} \yy }^2 ~.
\end{align}
Since solving \eqref{eq:ml_pos} still requires a 3D search, 
we propose the first subroutine, which provides an initial coarse position estimate with reduced complexity (2D search and 1D search), followed by an extremely efficient refinement with linear approximation. 
\subsubsection{Coarse Estimate}

We first assume that the \ac{UE} is static (i.e., $\vv=\boldzero\unit{m/s}$), then the velocity effects in the \ac{RIS} response model of~\eqref{eq:nf_steer}, as well as the dependency on the time index $\ell$ disappear. Hence, the simplified response model becomes
\begin{align} 
    [\aa(\pp)]_m &= \exp\left(-j \frac{2\pi}{\lambda}(\dm-\dr)\right) ~.
     \label{eq:a_vz}
\end{align}
Accordingly, we define
$\bW \triangleq [ \ww_1 \ \cdots \ \ww_L ] \in \complexset{M}{L}$
and 
$\hh(\pp) \triangleq \hh(\pp, \boldzero) = \bW^\top \aa(\pp)$, then the ML estimator in \eqref{eq:ml_pos} can now be expressed as
\begin{align}\label{eq:ml_pos_vel_vz}
		\pphat = \arg \min_{\pp}  \norm{\projnull{\bW^\top \aa(\pp)} \yy }^2.
\end{align}
To solve \eqref{eq:ml_pos_vel_vz}, we 
utilize a first-order Taylor  expansion to approximate the generic \ac{NF} \ac{RIS} response in \eqref{eq:a_vz} by its \ac{FF} version as [See Appendix~\ref{app:FF_approximation}]\footnote{Note that the \ac{FF} model only serves as an initial step and is used solely in line 3 of the algorithm. In contrast, all the other steps in all the subroutines use the \ac{NF} model.}
\begin{align} \label{eq:a_ff}
     [\aa(\pp)]_m & \approx [\aa(\theta, \phi)]_{m} =
    \exp{
    (- j (\pp_{m} - \ppr)^{\top}\kk(\phi, \theta))
    } ~,
\end{align}
where 
$\kk(\phi,\theta) = -\frac{2\pi}{\lambda}[\sin{\phi}\cos{\theta} ~, \hspace{1mm}
\sin{\phi}\sin{\theta} ~, \hspace{1mm}
\cos{\phi}]^\top.$
%
Leveraging the approximated model in~\eqref{eq:a_ff}, we performed a 2D search over $\theta \in [0, \, 2\pi]$ in azimuth and $\phi \in [0, \, \pi/2]$ in elevation to find $\thetahat$ and $\phihat$, respectively: 
\begin{align}
(\thetahat,\phihat) = \argmin_{\theta,\phi}  \norm{\projnull{\bW^\top \aa(\theta,\phi) } \yy }^2 ~.
\label{eq:2D_ff}
\end{align}
Then, based on the estimated angles, a linear distance search is performed 
\begin{align}
\rhohat = \argmin_{\rho} \norm{\projnull{\bW^\top \aa(\pp(\rho, \thetahat,\phihat)) } \yy }^2
    \label{eq:1D_distance},
\end{align}
to find $\rhohat$, using the \ac{NF} model~\eqref{eq:a_vz}. 
The steps \eqref{eq:2D_ff}--\eqref{eq:1D_distance} were proposed in \cite{abu2020near}. We extend the method by iteratively refining the 2D angle estimate with the \ac{NF} model
\begin{align}
(\thetahat,\phihat) = \argmin_{\theta,\phi} \norm{\projnull{\bW^\top \aa(\pp(\rhohat, \theta,\phi) } \yy }^2,
\label{eq:2D_nf}
\end{align}
and the distance estimate using \eqref{eq:1D_distance}. 
The entire procedure is summarized in Algorithm~\ref{alg:init_pos}.

\begin{algorithm}[t]
	\caption{Initialize Position and Gain Estimate: $(\pphat, \alphahat) = \initposgain(\yy)$}
	\begin{algorithmic}[1]
	    \State \textbf{Input:} Observation $\yy$ in \eqref{eq:yy}.
	    \State \textbf{Output:} Initial position estimate $\pphat$, initial gain estimate $\alphahat$.
	    \State Find initial $\theta$ and $\phi$ estimates 
     by solving \eqref{eq:2D_ff}.
	\State \textbf{While} \eqref{eq:1D_distance} and \eqref{eq:2D_nf} do not converge
	     \Indent
	        \State Update the distance estimate via \eqref{eq:1D_distance}.
	        \State Update the azimuth and elevation estimate via \eqref{eq:2D_nf}.
	     \EndIndent
	     \State \textbf{end while}
	     \State Compute the initial position estimate $\pphat$.
	     \State Compute the initial gain estimate $\alphahat$ via \eqref{eq:alphahat_update}.
	\end{algorithmic}
	\normalsize
 \label{alg:init_pos}
\end{algorithm}

\subsubsection{Refinement via Linearizations}\label{subsec:lin_init}
We now present the second subroutine, which is a local refinement to refine the coarse position estimate $\pz$ and account for the fact that the velocity is in reality non-zero. Two consecutive linearizations are applied to refine the position estimate: \textit{(i)} the linearization of the argument of the exponential in \eqref{eq:nf_steer}, i.e., $\flm (\pp, \vv)$ in \eqref{eq:flm}, and \textit{(ii)} the \ac{SAA} for the exponential itself.

\paragraph{Linearization of Exponential Argument}
We propose to linearize the phase term in the initial steering vector \eqref{eq:nf_steer} around $\pz$ to obtain a first-order approximation to the argument of the exponential in \eqref{eq:nf_steer}.
The  linearized \ac{RIS} response model can now be expressed as 
\begin{align}\label{eq:nf_steer_appr}
[\aa(\pp_\ell)]_m & \approx [\aat_{\ell}(\pp,\vv)]_m \triangleq
 \exp\left(-j \frac{2\pi}{\lambda} \Big\{ \flm(\pz, \vv) \right. \nonumber\\
 & \left.+ (\pp - \pz)^\top \gradF(\pp, \vv) \lvert_{\pp = \pz} \Big\} \right) ~,
\end{align}
where $\aat_{\ell}(\pp,\vv)$ contains the linearization of the exponential argument of  $\aa(\pp_\ell)$ (i.e., $\flm(\pp,\vv)$ in~\eqref{eq:flm} ) around $\pz$, and the gradient $\gradF(\pp, \vv)$ is derived in Appendix~\ref{app:gradient}.
Since the right-hand side of \eqref{eq:nf_steer_appr} depends on $\pp$
only via $\pd \triangleq \pp - \pz$, 
we can change the unknown from $\pp$ to $\pd$, which results in
%
%
\begin{align}
 [\aat_{\ell}(\pz+\pd,\vv)]_m &= \exp\left(j  [ \bbl + \pd^\top \ccl ] \right)~,\label{eq:nf_steer_appr3} \\
\bbl &\triangleq -\frac{2\pi}{\lambda}\flm(\pz, \vv) ~,   \label{eq:blm} 
    \\ 
\ccl &\triangleq -\frac{2\pi}{\lambda} \gradF(\pp, \vv) \lvert_{\pp = \pz} \in \realset{3}{1}.
    \label{eq:clm}
\end{align}
Here, the dependence of $\bbl$ and $\ccl$ on $\vv$ is dropped for the sake of notation convenience (and because $\vv$ is known).
%
%
%
%
\paragraph{Small Angle Approximation (SAA)}
We now invoke \ac{SAA} for \eqref{eq:nf_steer_appr3} to perform the second linearization phase. Since $\pd$ is expected to be small, we can linearize~\eqref{eq:nf_steer_appr3} around it by  invoking \ac{SAA} to obtain
\begin{align}
    [\aat_{\ell}(\pz+\pd,\vv)]_m 
    &\approx \exp(j  \bbl ) (1 + j \pd^\top \ccl) ~,
    \label{eq:nf_steer_appr_saa}
\end{align}
Based on the approximation in \eqref{eq:nf_steer_appr_saa}, we can compute $h_{\ell}(\pp,\vv)$ in \eqref{eq:hl_pv} as
\begin{align}
   & h_{\ell}(\pz+\pd,\vv) = \wwl^\top \aat_{\ell}(\pz+\pd,\vv)\\
   &= \sum_{m=1}^{M} [\wwl]_m [\aat_{\ell}(\pz+\pd,\vv)]_m\\
   & \approx \etal + j \pd^\top \bxil ~.
\end{align}
where $\etal \triangleq \sum_{m=1}^{M} [\wwl]_m \exp(j  \bbl )$ and  $\bxil \triangleq \sum_{m=1}^{M} [\wwl]_m \exp(j  \bbl ) \ccl$. 
%
%
Introducing $\etab \triangleq [\eta_1 \, \cdots \, \eta_L]^\top \in \complexset{L}{1} $ and  $\bxib \triangleq [\bxi_1 \, \cdots \, \bxi_L] \in \complexset{3}{L} $, 
$\hh(\pp=\pz+\pd, \vv)$ in \eqref{eq:hl_pv_vec} becomes
\begin{align} \label{eq:hh_appr}
    \hh(\pz+\pd, \vv) &\approx \etab + j \bxib^\top \pd ~,
\end{align}
%
which implies that 
the ML estimator \eqref{eq:ml_pos_vel_gain} can be approximated as 
\begin{align}\label{eq:ml_pos_gain_appr}
		(\alphahat, \pdhat) = \arg \min_{\alpha, \pd}  \norm{\yy - \alpha \,  (\etab + j \bxib^\top \pd )  }^2 ~,
\end{align}
where we recall that we estimate the residual position $\pd$ instead of the actual position $\pp$.
We propose to solve \eqref{eq:ml_pos_gain_appr} via alternating updates of $\alpha$ and $\pd$, yielding (see Appendix~\ref{app:pd})
%
%
\begin{align} \label{eq:pdhat}
    \pdhat &  = \frac{1}{\abs{\alphahat}^2} \big( \realp{\bxib^\conj \bxib^\top} \big)^{-1} \imp{ \bxib \big(    \abs{\alphahat}^2 \etab^\conj   -  \alphahat \, \yy^\conj  \big)} ~, \\
     \alphahat&  = \frac{ (\etab + j \bxib^\top \pdhat )^\h \yy }{ \norm{\etab + j \bxib^\top \pdhat}^2 }.\label{eq:alphahat_p}
\end{align}
%
%
Algorithm~\ref{alg:ref_pos} summarizes the above procedure.

\begin{algorithm}[t]
	\caption{Refine Position and Gain Estimate with Fixed Velocity: $(\pphat, \alphahat) = \reposgain(\yy, \vv, \pz, \alpha_0)$}
	\begin{algorithmic}[1]
	    \State \textbf{Input:} Observation $\yy$ in \eqref{eq:yy}, velocity $\vv$, initial position estimate $\pz$, and initial gain estimate $\alpha_0$.
	    \State \textbf{Output:} Refined position estimate $\pphat$, refined gain estimate $\alphahat$.
	    \State Set $\alphahat = \alpha_0$.
	    \State \textbf{while} the objective in \eqref{eq:ml_pos_gain_appr} does not converge
	     \Indent
            \State Compute $\etab$ and $\bxib$ in \eqref{eq:ml_pos_gain_appr} using $\vv$ and $\pz$.
	    \State Update the residual position estimate $\pdhat$ via \eqref{eq:pdhat}.
	        \State Update the gain estimate $\alphahat$ via \eqref{eq:alphahat_p}.
	     \EndIndent
	     \State \textbf{end while}
	     \State Refine the position estimate by setting $\pphat = \pz + \pdhat$.
	\end{algorithmic}
	\normalsize
 \label{alg:ref_pos}
\end{algorithm}

\subsection{Velocity Estimate with Known Position}
\label{subsec:VelEstimate}

In this section, we present an iterative refinement routine, in closed form, to estimate the velocity of the \ac{UE} assuming knowledge over its position. The procedure is similar to Section~\ref{subsec:PosEstimate} but operating on residual velocity rather than residual position. 
However, unlike  the position refinement subroutine, the linearization of the argument of the exponential, as done in the first step of Section~\ref{subsec:PosEstimate},  is not needed here, since the velocity already appears linearly in the argument of the exponential in the steering vector. This means that the only approximation needed is the \ac{SAA}, which can be around any  initial value, including $\vv=\boldsymbol{0}$, provided the velocity error is low or moderate.


We denote by $\vz$ the initial velocity estimate and $\vd = \vv - \vz$ the velocity residual resulting from estimation inaccuracies. The RIS response \eqref{eq:nf_steer} can be expressed as 
\begin{align}
    [\aa(\pp_\ell)]_{m} 
    = e^{j\beta_{m}}
    \exp \big( j  \vv^{\top} \gammabm \ell\Ts \big) ,
\end{align}
where
    $\beta_m = -\frac{2\pi}{\lambda}
    \big( \norm{\pp_{m} - \pp} 
    -
    \norm{\ppr - \pp}\big)$ and
    $\gammabm = -\frac{2\pi}{\lambda}
    \big(\uu_{m}(\pp) \big)$.
Note that the dependence of $\beta_m$ and $\gammabm$ on $\pp$ is dropped for notational convenience because $\pp$ is given. 
We then express the latter in terms of the residual velocity $\vd$ as
\begin{align}
    [\aa(\pp_\ell)]_{m}
    &= \exp\big(j\beta_{m}\big)
    \exp \big( j\vz^{\top}\gammabm \ell\Ts\big)
    \exp \big( j\vd^{\top}\gammabm \ell\Ts\big)~,\label{eq:b(v)_approx}
\end{align} 
and linearize around $\vd$, which we expect to be small, by employing the \ac{SAA}
\begin{align}
    [\aa(\pp_\ell)]_{m} \approx 
    \exp\big(j\beta_{m}\big)\exp \big( j\vz^{\top}\gammabm \ell\Ts\big)
    \big(1 + j\vd^{\top} \gammabm \ell\Ts\big) ~. \label{eq:a_approx}
\end{align}
Based on this approximation we then compute the terms in \eqref{eq:hl_pv} reflecting \eqref{eq:a_approx} as
$ [\hl(\pp,\vz+\vd)]_m = \sum_{m=1}^{M} [\wwl]_m [\aa(\pp_\ell)]_m ~
\approx \nul + j \vd^\top\mubl
$
where
$\nul = \sum_{m=1}^{M} [\wwl]_m\exp\big(j\beta_{m}\big)\exp \big( j\vz^{\top}\gammabm \ell\Ts\big),
$
    and
    $\mubl = \sum_{m=1}^{M} 
    [\wwl]_m
    \exp\big(j\beta_{m}\big)\exp \big( j\vz^{\top}\gammabm \ell\Ts\big)
     \gammabm \ell\Ts.
$
Using the definition in~\eqref{eq:hl_pv_vec}, the vector form of $\hl(\pp,\vz+\vd)$ can be expressed as
\begin{align}
    \hh(\pp,\vz+\vd) &\approx \nub + j \Mb^\top\vd ~,\label{eq:h_resid}
\end{align}
where  $\nub \triangleq [\nu_1 \ \dots \ \nu_L]^\top$ and $\Mb \triangleq [\vect{\mu}_1 \ \dots \ \vect{\mu}_L]^\top \in \mathbb{C}^{3\times L}$.
The \ac{ML} estimation for the residual velocity from \eqref{eq:ml_pos_vel_gain} can be expressed as
\begin{align}\label{eq:ml_vel_gain_appr}
		(\alphahat, \vdhat) = \arg \min_{\alpha, \vd}  \norm{\yy - \alpha \,  (\nub + j \Mb^\top \vd)  }^2 ~,
\end{align}
We propose to solve \eqref{eq:ml_vel_gain_appr} via alternating updates of $\alpha$ and $\vd$, yielding
\begin{align} \label{eq:vdhat}
    \vdhat & = \frac{1}{\abs{\alphahat}^2} \big( \realp{\Mb^\conj \Mb^\top} \big)^{-1} \imp{ \Mb \big(    \abs{\alphahat}^2 \nub^\conj   -  \alphahat \, \yy^\conj  \big)} ,\\
      \alphahat & = \frac{ (\nub + j \Mb^\top \vdhat )^\her \yy }{ \norm{\nub + j \Mb^\top \vdhat}^2 } ~.\label{eq:alphahat_v}
\end{align}
Note that here as well, the mild condition of $L \geq 3$ transmissions is required to compute the closed form \ac{UE} velocity using \eqref{eq:vdhat}.
Algorithm~\ref{alg:ref_vel} summarizes the procedure.

\begin{algorithm}[t]
	\caption{Refine Velocity Estimate with Fixed Position and Gain: $\vvhat = \revel(\yy, \vz, \pp, \alpha)$}
	\begin{algorithmic}[1]
	    \State \textbf{Input:} Observation $\yy$ in \eqref{eq:yy}, position $\pp$, initial velocity estimate $\vz$, and gain $\alpha_0$.
	    \State \textbf{Output:} Refined velocity estimate $\vvhat$, refined gain estimate $\alphahat$.
	    \State Set $\alphahat = \alpha_0$.
	    \State \textbf{while} the objective in \eqref{eq:ml_vel_gain_appr} does not converge
	     \Indent
	   \State Compute $\nub$ and $\Mb$ in \eqref{eq:ml_vel_gain_appr} using $\pp$ and $\vz$.
            \State Update the residual velocity estimate $\vdhat$ via \eqref{eq:vdhat}.
	\State Update the gain estimate $\alphahat$ via \eqref{eq:alphahat_v}.
	     \EndIndent
	\State \textbf{end while}
	\State Refine the velocity estimate by setting $\vvhat = \vz + \vdhat$.
	\end{algorithmic}
	\normalsize
        \label{alg:ref_vel}
\end{algorithm}


\subsection{Time Complexity Analysis}
\label{sec:comp}
We now present the complexity scaling of the algorithms as a function of the \ac{RIS} size $M$ and number of pilot transmissions $L$, as well as the number of iterations in Algorithm~\ref{alg:overall},  Algorithm~\ref{alg:init_pos}, Algorithm~\ref{alg:ref_pos}, and Algorithm~\ref{alg:ref_vel}, denoted by $I_1$, $I_2$, $I_3$, and $I_4$, respectively. We denote the complexity of Algorithm $i$ by $\mathcal{C}_i$. 

We find that $\mathcal{C}_3=\mathcal{O}(I_3 L M)$ and $\mathcal{C}_4=\mathcal{O}(I_4 L M)$, mainly due to the computation of line 5 (i.e., $\etab$, $\bxib$,   $\nub$, and $\Mb$) in both methods.  Moving this line before the while loop will turn out to not severely affect performance, but leads to a significant complexity reduction $\mathcal{C}_3=\mathcal{O}(I_3 L)$ and $\mathcal{C}_4=\mathcal{O}(I_4 L)$. 
It is also readily verified that $C_2=\mathcal{O}(L M I_2 \max\{K_\theta \times K_\phi, K_\rho\})$, where  $K_\theta$ (resp.~$K_\phi$ and $K_\rho$) denotes the grid size for the search over $\theta$ (resp.~$\phi$, $\rho$). 

Finally, introducing $I_{\text{GD}}$ as the number of gradient descent iterations,  we find that 
\begin{align}
\mathcal{C}_1& =\mathcal{C}_2 + I_1 (\mathcal{C}_3+\mathcal{C}_4) + \mathcal{O}(I_{\text{GD}}LM)\\
    & =\mathcal{O}(L M I_2 \max\{K_\theta \times K_\phi, K_\rho\}) + \mathcal{O}(I_1 I_3 L) \notag \\
    &+ \mathcal{O}(I_1 I_4 L) + \mathcal{O}(I_{\text{GD}}LM),
\end{align}
provided line 5 is moved before the while loop in Algorithms ~\ref{alg:ref_pos}--\ref{alg:ref_vel}. 

\section{Numerical Simulations and Results}
\label{sec:Simulations}

In order to evaluate the effectiveness of the proposed estimation framework, numerical simulations were conducted in a canonical indoor scenario.\footnote{In the numerical simulations, for the generative model as well as the bounds evaluation, we use the approximated expression in~\eqref{eq:flm}.}
\subsection{Scenario Definition and Performance Metrics}\label{sec:scen_def_and_perf_met}
In the following, we consider random \ac{RIS} phase profile design and we list, in Table~\ref{table:params}, the scenario parameters~\cite{Rahal_Localization-Optimal_RIS}.
%
We evaluate the performance of the algorithmic variants introduced in section~\ref{sec:proposed_methods} (under different prior knowledge assumptions), while comparing them with both theoretical bounds (derived in Appendix~\ref{app:PEB}) and that of a \ac{SoTA} algorithm assuming zero velocity from \cite{abu2020near}. 
Hence, such a comparison brings forward the advantage of taking the extra phase shifts induced by velocity (\ac{UE}'s mobility) into consideration.
In the following, we refer to the algorithms as follows: 
\begin{itemize}
    \item \emph{\Ac{GS}:} This method refers to Algorithm~\ref{alg:init_pos}, providing a position estimate assuming a zero velocity.
    \item \emph{\ac{CF} Position Refinement:} This method refers to the closed-form position refinement routine of Algorithm~\ref{alg:ref_pos}.
    \item \emph{\ac{CF} Velocity Refinement:} This method refers to the closed-form velocity refinement procedure of Algorithm~\ref{alg:ref_vel}.
    \item \emph{Global Refinement:} This methods is the 6D gradient descent  in Algorithm~\ref{alg:overall}, line 9, which is used to enhance the accuracy of both position and velocity estimation.
\end{itemize} 

We also assess the sensitivity of the proposed approach to critical system parameters, such as the distance to the \ac{RIS}, the norm of the velocity vector, or small-scale fading effects resulting from uncontrolled multipath components (i.e., besides the used \ac{RIS}-reflected path). 
In terms of geometry, the \ac{RIS} is located at the \emph{xy}-plane center and the \ac{UE}'s position is defined as 
$\pp = \rho \vect{i}$ where $\rho \in [1,10]\unit{m}$ denotes the \ac{RIS}-\ac{UE} distance and $\vect{i} = \nicefrac{[-1,2,1]^\top}{\norm{[-1,2,1]^\top}}$ acts as a unit vector. Similarly, the \ac{UE}'s velocity is defined as $\vv = v \vect{i}$ where $v \in [0, 20]\unit{m/s}$ denotes the speed.\footnote{The velocity vector utilized in the simulations is radial w.r.t.~the \ac{RIS}. We have verified that the proposed algorithms operate close to the corresponding bounds, for other velocity vectors as well.} According to the approximation in Appendix~\ref{app:flm}, the \ac{UE} speed $v$ should be below the limit computed via~\eqref{eq:v_limit}, given the parameters in Table~\ref{table:params}. More precisely, a minimum distance condition $\norm{\bm{\delta}_m} > 1 \, \rm{m}, \,   \forall m$ is considered, as well as maximum values for $L = 50$ and $\Ts = 100\unit{\mu s}$. Under these settings, the condition $\norm{\vv} L\Ts \ll \norm{\bm{\delta}_m}, \forall m$ simplifies to $\norm{\vv} \ll 200 \, \rm{m/s}$, a condition that aligns with our simulations as we consider speeds up to $20 \, \rm{m/s}$. 

Performance is  evaluated through the \ac{RMSE} of the position and velocity, which is calculated over 1000 random measurement noise realizations for each tested \ac{UE} position-velocity configuration. The \acp{RMSE} are compared to the corresponding error bounds, \ac{PEB} and \ac{VEB}.

We found that in Algorithms~\ref{alg:ref_pos} and~\ref{alg:ref_vel}, when the linearization in line 5  is only executed once before the loop instead of every iteration, the performance is nearly identical to performing the linearization every iteration. Given the complexity analysis in Section \ref{sec:comp}, from now on we only use this low-complexity version the \ac{CF} refinement methods. 
\begin{table}
\centering
\resizebox{\columnwidth}{!} {
\begin{tabular}{ |c|c||c|c| } 
 \hline
 parameter & value & parameter & value\\
 \hline
 Frequency & $f_{c} = 28\unit{GHz}$ & Wavelength & $\lambda \approx 1.07\unit{cm}$\\
 Bandwidth & $W = 1\unit{MHz}$ & Power transmitted & $P = 20\unit{dBm}$\\ 
 Noise PSD & $N_0 = -174\unit{dBm/Hz}$ & Noise figure & $n_f = 8\unit{dB}$\\ 
  RIS loc. & $\ppr = [0, 0, 0]^\top\unit{m}$ & BS loc. & $\ppb = [3, 3, 1]^\top\unit{m}$\\ 
  RIS size & $M = 32\times 32$ elements &  Transmissions & $L = 40$ \\
 \hline
\end{tabular}\vspace{-4mm}
}
\caption{General simulation parameters.}
    \label{table:params}
\end{table}

\subsection{Performance of 3D Estimation}
We first discuss the performance of the 3D estimation algorithms for position and velocity separately.
\subsubsection{Position Estimation with Known Velocity}
%
\begin{figure}
     \centering
     \resizebox{1\columnwidth}{!}{
%
%
\definecolor{amber}{rgb}{1.00000,0.75,0.00000}%
\begin{tikzpicture}

\begin{axis}[%
width=4.521in,
height=2.566in,
at={(0.758in,0.481in)},
scale only axis,
xmin=1,
xmax=10,
xlabel style={font=\large, font = \color{white!15!black}},
xlabel={RIS-UE distance$\unit{[m]}$},
ymode=log,
ymin=0.001,
ymax=1,
yminorticks=true,
ylabel style={font=\large, font =\color{white!15!black}},
ylabel={RMSE$\unit{[m]}$},
axis background/.style={fill=white},
xmajorgrids,
ymajorgrids,
yminorgrids,
legend style={font = \large, at={(0.97,0.03)}, anchor=south east, legend cell align=left, align=left, draw=white!15!black}
]
\addplot [color=blue, line width=1.8pt, mark=o, mark options={solid, blue}, mark size=5.0pt]
  table[row sep=crcr]{%
1	0.000129950238847488\\
2	0.000732228522534494\\
3	0.00201524747153256\\
4	0.00413172275623937\\
5	0.00720643734053307\\
6	0.0113452026962657\\
7	0.0166391726608599\\
8	0.0231673114427887\\
9	0.0309979747579604\\
10	0.0401900172310992\\
};
\addlegendentry{PEB}

\addplot [color=red, line width=2pt, mark=asterisk, mark size=5.0pt, mark options={solid, red}]
  table[row sep=crcr]{%
1	0.00772158154657697\\
2	0.0106814823602189\\
3	0.0185222475670745\\
4	0.0289107357873369\\
5	0.0419991974479198\\
6	0.0544582364693642\\
7	0.0788293359025073\\
8	0.0948084222780279\\
9	0.109060952322833\\
10	0.146922392452558\\
};
\addlegendentry{Grid Search}

\addplot [color=amber, line width=2pt, mark=square, mark size=5.0pt, mark options={solid, amber}]
  table[row sep=crcr]{%
1	0.000177389027258634\\
2	0.000799906066096256\\
3	0.00217255345259909\\
4	0.0043701453545206\\
5	0.00801592057533089\\
6	0.0117617671052578\\
7	0.0174555527076974\\
8	0.0260286836268159\\
9	0.0298482820653515\\
10	0.0429749552934417\\
};
\addlegendentry{CF Position Refinement}

\addplot [color=green, line width=1.8pt, dashed, mark=asterisk, mark options={solid, green}, mark size=4.0pt,]
  table[row sep=crcr]{%
1	 0.0378\\
2	 0.0580\\
3	 0.1126\\
4	 0.2073\\
5	 0.3098\\
6	 0.4394\\
7	 0.6351\\
8	 0.7618\\
9	 1.0455\\
10	 1.3261\\
};
\addlegendentry{Method from \cite{abu2020near}}

\end{axis}

\end{tikzpicture}%
}
    \caption{\ac{RMSE} of position estimation with known \ac{UE} velocity ($v = 1\unit{m/s}$) using Algorithms~\ref{alg:init_pos} and~\ref{alg:ref_pos}, along with the corresponding \ac{PEB}, as a function of the \ac{RIS}-\ac{UE} distance.}
    \label{fig:alg2}
\end{figure}
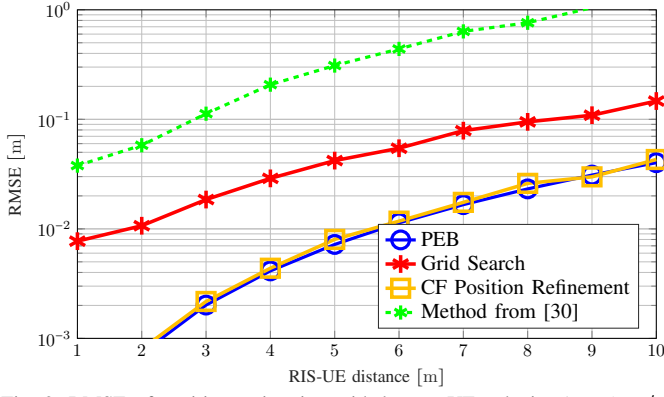
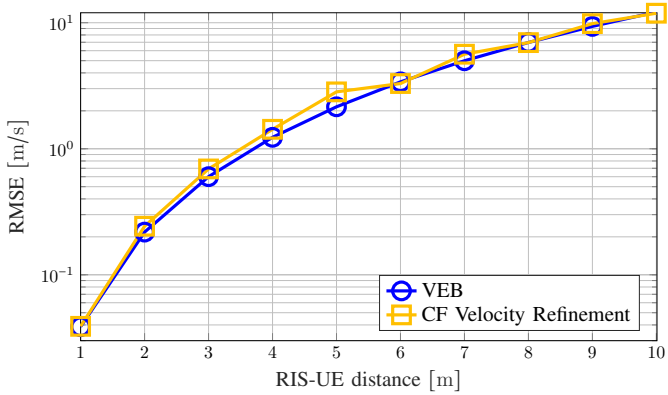
\begin{figure}
     \centering
     \resizebox{1\columnwidth}{!}{
%
%
\definecolor{amber}{rgb}{1.00000,0.75,0.00000}%
\begin{tikzpicture}

\begin{axis}[%
width=4.521in,
height=2.572in,
at={(0.758in,0.482in)},
scale only axis,
xmin=1,
xmax=10.01,
xlabel style={font= \large  \color{white!15!black}},
xlabel={RIS-UE distance$\unit{[m]}$},
ymode=log,
ymin=0.03,
ymax=12,
yminorticks=true,
ylabel style={font=\large \color{white!15!black}},
ylabel={RMSE$\unit{[m/s]}$},
axis background/.style={fill=white},
xmajorgrids,
ymajorgrids,
yminorgrids,
legend style={font = \large, at={(0.97,0.03)}, anchor=south east, legend cell align=left, align=left, draw=white!15!black}
]
\addplot [color=blue, line width=1.8pt, mark=o, mark size=5.0pt, mark options={solid, blue}]
  table[row sep=crcr]{%
1	0.0387185909880596\\
2	0.218097554937013\\
3	0.600530661291385\\
4	1.2324279510458\\
5	2.15268642798643\\
6	3.39549870478391\\
7	4.99174347864151\\
8	6.9697990769535\\
9	9.35606924560707\\
10	12.175346593478\\
};
\addlegendentry{VEB}

\addplot [color=amber, line width=1.8pt, mark=square, mark options={solid, amber}, mark size=5.0pt]
  table[row sep=crcr]{%
1	0.0389901116952202\\
2	0.242765923739332\\
3	0.691570646970451\\
4	1.42659905238129\\
5	2.82771794751217\\
6	3.29262929398094\\
7	5.62644131750299\\
8	6.95588456396754\\
9	9.83697865502577\\
10	11.8998100471153\\
};
\addlegendentry{CF Velocity Refinement}


\end{axis}
\end{tikzpicture}%
}
    \caption{\ac{RMSE} of velocity estimation with known position and an unkown velocity ($v = 1\unit{m/s}$) (i.e., Algorithm~\ref{alg:ref_vel}) and corresponding \ac{VEB} as a function of \ac{RIS}-\ac{UE} distance.}
    \label{fig:alg3}
\end{figure}

First, in Fig.~\ref{fig:alg2}, we show the \ac{RMSE} of the position estimation for the two algorithmic steps of Section~\ref{subsec:PosEstimate}, along with the corresponding \ac{PEB}, as a function of the \ac{RIS}-\ac{UE} distance and with a fixed velocity of $v = 1\unit{m/s}$. One can notice that the \ac{PEB} (in blue) remains at a sub-cm level at short RIS-UE distances (i.e., below $6\unit{m}$) but then increases up to a few cm as the UE moves away from the RIS. The first part of our estimation routine, which relies on \ac{GS} (in red) routine in~\ref{alg:init_pos}, 
apparently suffers from a big performance gap compared to the bound, this is caused by the multiple approximations made during the development of this algorithm (See Section~\ref{subsec:PosEstimate}). However, feeding the output of this initial search routine into the \ac{CF} position refinement routine (in amber), i.e., Algorithm~\ref{alg:ref_pos} enables to reach the performance bound, whatever the distance. 
To better assess the performance of our algorithm, we have also benchmarked our algorithm with the closest contribution from the \ac{SoTA}. More specifically, in~\cite{abu2020near}, the authors devised a grid-search-based algorithm to estimate the position of the \ac{UE} by exploiting the signal’s wavefront curvature. This approach accounts for the effects of geometric \ac{NF} on positioning while assuming no velocity in the model. 
The method from \cite{abu2020near} involves
three distinct 1D search procedures over the three spherical dimensions. For this purpose, they performed a Jacobi Anger expansion to be able to separate both angles in the \ac{RIS} model. Fig.~\ref{fig:alg2} shows that the method from \cite{abu2020near} leads to a significant performance degradation. 
\subsubsection{Velocity Estimation with Known Position}
In Fig.~\ref{fig:alg3}, we show the \ac{RMSE} of velocity estimation for the refinement algorithm of Section~\ref{subsec:VelEstimate}, along with the corresponding \ac{VEB}, still as a function of the \ac{RIS}-\ac{UE} distance. Without loss of generality, the actual velocity value was set to $1\unit{m/s}$ in this example, but further studies include higher values to evaluate the sensitivity to more extreme \ac{UE} mobility (see Section~\ref{subsec:Sensitivity_to_Velocity}). Note that the tested algorithm does not include an extensive grid search like before, but uniquely the \ac{CF} velocity refinement routine, which is called with an initial velocity value of $0\unit{m/s}$. We thus observe 
that the \ac{RMSE} of this estimation almost coincides with the \ac{VEB} curve, regardless of the \ac{RIS}-\ac{UE} distance. 

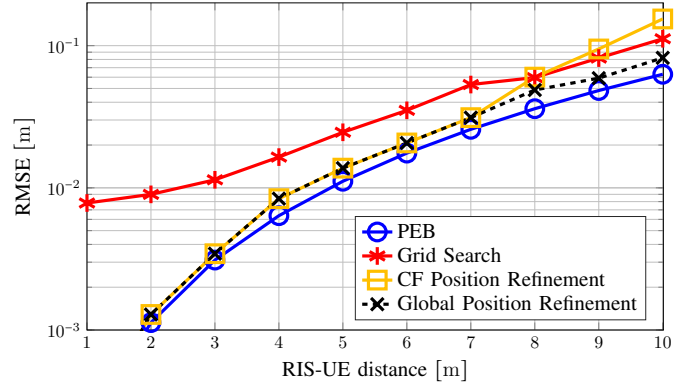
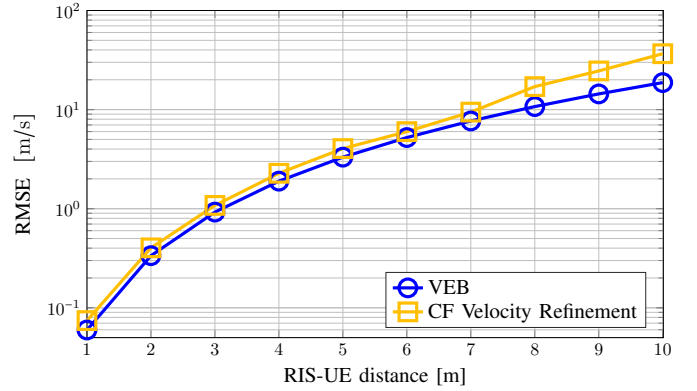
\begin{figure}
     \centering
     \label{fig:err_vs_dist_v1}
     \begin{subfigure}[b]{1\columnwidth}
         \centering
         \resizebox{1\columnwidth}{!}{
%
%
\definecolor{amber}{rgb}{1.00000,0.75,0.00000}%
\begin{tikzpicture}

\begin{axis}[%
width=4.521in,
height=2.566in,
at={(0.758in,0.481in)},
scale only axis,
xmin=1,
xmax=10.01,
xlabel style={font=\color{white!15!black}, font = \large},
xlabel={RIS-UE distance$\unit{[m]}$},
ymode=log,
ymin=0.001,
ymax=0.2,
yminorticks=true,
ylabel style={font=\color{white!15!black}, font = \large},
ylabel=RMSE$\unit{[m]}$,
axis background/.style={fill=white},
xmajorgrids,
ymajorgrids,
yminorgrids,
legend style={at={(0.97,0.03)}, anchor=south east, legend cell align=left, align=left, draw=white!15!black, font = \large}
]
\addplot [color=blue, line width=1.8pt, mark=o, mark size=5.0pt, mark options={solid, blue}]
  table[row sep=crcr]{%
1	0.000200309548899163\\
2	0.00112746417791503\\
3	0.00310406244337275\\
4	0.00636999002870104\\
5	0.0111262925794762\\
6	0.0175496982655204\\
7	0.025799806027183\\
8	0.03602329094886\\
9	0.0483566220518732\\
10	0.0629279408346325\\
};
\addlegendentry{PEB}

\addplot [color=red, line width=1.8pt, mark=asterisk, mark options={solid, red}, mark size=5.0pt,]
  table[row sep=crcr]{%
1	0.00782069383618277\\
2	0.00899698317923266\\
3	0.0113913147155294\\
4	0.0164543883729687\\
5	0.0245858010352725\\
6	0.0350841204281149\\
7	0.0532814006751122\\
8	0.0596429250910835\\
9	0.0818091929789828\\
10	0.111856079854129\\
};
\addlegendentry{Grid Search}


\addplot [color=amber, line width=1.8pt, mark=square, mark options={solid, amber}, mark size=5.0pt]
  table[row sep=crcr]{%
1	0.000217398702930395\\
2	0.00128612709147153\\
3	0.0034407851874406\\
4	0.00839154946104195\\
5	0.0137571038841984\\
6	0.0206879651626686\\
7	0.0311718361432946\\
8	0.0600851471928036\\
9	0.0947955914679849\\
10	0.154116585511851\\
};
\addlegendentry{CF Position Refinement}


\addplot [color=black, dashed, line width=1.8pt, mark=x, mark options={solid, black}, mark size=5.0pt,]
  table[row sep=crcr]{%
1	0.000217485968149497\\
2	0.00128610559022379\\
3	0.00344181042123817\\
4	0.00839242789154545\\
5	0.013764897604342\\
6	0.0206761203327215\\
7	0.0311642064754117\\
8	0.0487981382834974\\
9	0.0590791073588498\\
10	0.0821366385683579\\
};
\addlegendentry{Global Position Refinement}


\end{axis}
\end{tikzpicture}%
}
         \caption{\ac{RMSE} of position estimation and related \ac{PEB} versus \ac{RIS}-\ac{UE} distance.}
        \label{fig:p_err_vs_dist_v1}
    \end{subfigure}
     \hfill
     \vspace{0.5mm}
     \begin{subfigure}[b]{1\columnwidth}
         \centering
         \resizebox{1\columnwidth}{!}{
%
%
\definecolor{amber}{rgb}{1.00000,0.75,0.00000}%
\begin{tikzpicture}

\begin{axis}[%
width=4.521in,
height=2.566in,
at={(0.758in,0.481in)},
scale only axis,
xmin=1,
xmax=10.01,
xlabel style={font=\color{white!15!black}, font = \large},
xlabel={RIS-UE distance [m]},
ymode=log,
ymin=0.05,
ymax=100,
yminorticks=true,
ylabel style={font=\color{white!15!black}, font = \large},
ylabel=RMSE $~\mathrm{[m/s]}$,
axis background/.style={fill=white},
xmajorgrids,
ymajorgrids,
yminorgrids,
legend style={at={(0.97,0.03)}, anchor=south east, legend cell align=left, align=left, draw=white!15!black, font = \large}
]
\addplot [color=blue, line width=1.8pt, mark=o, mark size=5.0pt, mark options={solid, blue}]
  table[row sep=crcr]{%
1	0.0597044533957826\\
2	0.335749217465892\\
3	0.924217940539338\\
4	1.89653221394332\\
5	3.31254797420148\\
6	5.22488072740178\\
7	7.68104601287871\\
8	10.7247115664693\\
9	14.3965065213243\\
10	18.7345807230352\\
};
\addlegendentry{VEB}

\addplot [color=amber, line width=1.8pt, mark=square, mark options={solid, amber}, mark size=5.0pt,]
  table[row sep=crcr]{%
1	0.0742802796912528\\
2	0.402652451638548\\
3	1.07935748579632\\
4	2.27723370069387\\
5	4.04386521926453\\
6	5.99467398116186\\
7	9.43477812044036\\
8	17.061953523183\\
9	24.5600549664894\\
10	36.6497547666272\\
};
\addlegendentry{CF Velocity Refinement}



\end{axis}
\end{tikzpicture}%
}
         \caption{\ac{RMSE} of velocity estimation and related \ac{VEB} versus \ac{RIS}-\ac{UE} distance.}
        \label{fig:v_err_vs_dist_v1}
     \end{subfigure}
     \caption{6D estimation errors and related theoretical bounds versus \ac{RIS}-\ac{UE} distance, for a constant \ac{UE} velocity ($v = 1\unit{m/s}$).}
\end{figure}
\subsection{Performance of 6D estimation}
After simulating the 3D position and 3D velocity estimation algorithms separately, we now evaluate the performance of 6D position and velocity estimation with no prior knowledge regarding the \ac{UE} state. 
Figs.~\ref{fig:p_err_vs_dist_v1} and~\ref{fig:v_err_vs_dist_v1} first show the \ac{RMSE} of position and velocity estimation respectively, alongside their related theoretical bounds, versus the \ac{RIS}-\ac{UE} distance 
for a static \ac{UE}, i.e., with $v = 1\unit{m/s}$. 
The curves in Fig.~\ref{fig:p_err_vs_dist_v1} show that Algorithm~\ref{alg:init_pos} (\ac{GS} - in red) alone does not perform well compared to the \ac{PEB}, especially at short distance,  
whereas Algorithm~\ref{alg:ref_pos} (\ac{CF} Refinement routine - in amber) significantly boosts the estimation performance at short distances, and even touches the \ac{PEB} curve at short-to-mid ranges. However, as the \ac{RIS}-\ac{UE} distance increases, we notice a drop in performance, this is caused by implementing the approximation in~(\ref{eq:nf_steer_appr_saa}) whose error increases as the \ac{UE} moves further from the \ac{RIS}. 
For this reason, a Global Refinement (in dashed black) routine has been implemented to enhance estimation mostly at far distances, where the previous algorithms fail. We did not perform any 6D grid search to compare with our results since: \emph{i)} the final estimate, i.e., after all the refinement steps, satisfies the bounds, and \emph{ii)} it is computationally infeasible to search in over 6 dimensions.
Moreover, we have performed a quantitative complexity analysis and found that, 
as expected, the \ac{GS} subroutine is the most time-consuming, at around $85\%$ of the runtime, whereas the refinement subroutines take up less than $15\%$ of the total runtime.

\subsection{Sensitivity Analyses}
Now that we have established the performance of the proposed method and its ability to attain the \acp{CRLB}, we move on to study the sensitivity to the speed, to uncontrolled multipath, and to the \ac{SNR}. 
\subsubsection{Sensitivity to the Speed}\label{subsec:Sensitivity_to_Velocity}
The same simulation setup is now used to study the effect of the \ac{UE}'s velocity on our algorithms. 
Thus, we assume a dynamic \ac{UE}, with a velocity varying from $0\unit{m/s}$ to around $50\unit{m/s}$, 
while the \ac{RIS}-\ac{UE} distance is set to $2\unit{m}$. The corresponding results are displayed in Fig.~\ref{fig:p_err_vs_vel_d1}. We can hence see clearly the negative effect of large velocity on the \ac{GS} algorithm, while the rest of the other routines still continue performing well, in compliance with the \ac{PEB} behavior, suggesting that our overall estimation framework is fairly robust against high \ac{UE} mobility.
%
%
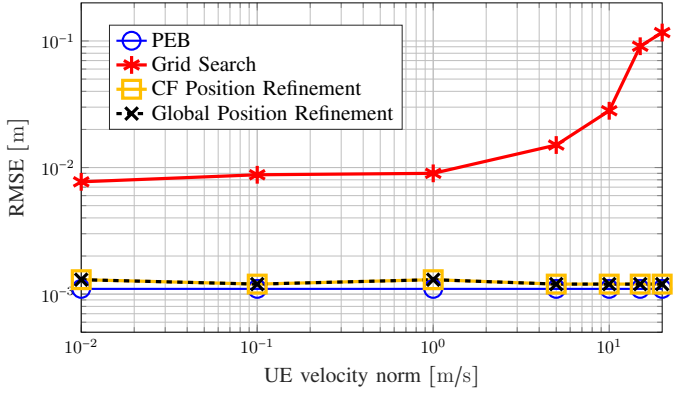
\begin{figure}
    \centering
    \resizebox{1\columnwidth}{!}{
%
\definecolor{amber}{rgb}{1.00000,0.75,0.00000}%

\begin{tikzpicture}

\begin{axis}[%
width=4.521in,
height=2.563in,
at={(0.758in,0.484in)},
scale only axis,
xmode=log,
xmin=0.01,
xmax=20,
xminorticks=true,
xlabel style={font=\large \color{white!15!black}},
xlabel={UE velocity norm$\unit{[m/s]}$},
ymode=log,
ymin=0.0005,
ymax=0.2,
yminorticks=true,
ylabel style={font=\large \color{white!15!black}},
ylabel={RMSE$\unit{[m]}$},
axis background/.style={fill=white},
xmajorgrids,
xminorgrids,
ymajorgrids,
yminorgrids,
legend style={font = \large, at={(0.048,0.615)}, anchor=south west, legend cell align=left, align=left, draw=white!15!black}
]
\addplot [color=blue, line width=1.2pt, mark=o, mark size = 5.0pt, mark options={solid, blue}]
  table[row sep=crcr]{%
0.01  0.0011\\
0.1	  0.0011\\
1	  0.0011\\
5	  0.0011\\
10	  0.0011\\
15	  0.0011\\
20	  0.0011\\
};
\addlegendentry{PEB}

\addplot [color=red, line width=1.8pt, mark=asterisk, mark size = 5.0pt, mark options={solid, red}]
  table[row sep=crcr]{%
0.01	0.00771242320506298\\
0.1	0.00876812981249091\\
1	0.00899910313745142\\
5	0.0150638370422127\\
10	0.0280981549811748\\
15 0.0908642192497560\\
20 0.116574300684297\\
};
\addlegendentry{Grid Search}

\addplot [color=amber, line width=1.8pt, mark=square, mark size = 5.0pt, mark options={solid, amber}]
  table[row sep=crcr]{%
0.01	0.0013\\
0.1	    0.0012\\
1	    0.0013\\
5	    0.0012\\
10	    0.0012\\
15      0.0012\\
20      0.0012\\	
}; 
\addlegendentry{CF Position Refinement}

\addplot [color=black, dashed, line width=1.8pt, mark=x, mark size = 5.0pt, mark options={solid, black}]
  table[row sep=crcr]{%
0.01	0.0013\\
0.1	    0.0012\\
1	    0.0013\\
5	    0.0012\\
10	    0.0012\\
15      0.0012\\
20      0.0012\\
};
\addlegendentry{Global Position Refinement}

\end{axis}
\end{tikzpicture}%
}
    \caption{\ac{UE}'s mobility effect on the position estimation and \ac{PEB} in \ac{NF}. The \ac{RIS}-\ac{UE} distance is set to $\rho = 2\unit{m}$.}    \label{fig:p_err_vs_vel_d1}
\end{figure}

\subsubsection{Sensitivity to Uncontrolled Multipath}

We now discuss the performance in the presence of a multipath-rich channel. The \ac{BS} is considered to be directive, and hence only one \ac{LoS} path exists in the \ac{BS}-\ac{RIS} channel~\cite{multi_abrardo2021,multi_jiang2023}. On the other hand, the channel between the \ac{RIS} and the \ac{UE} is modeled as Rician~\cite{multi_zhi2022,multi_peng2022}, which yields a change in the signal model of~(\ref{eq:obs_model}) into
\begin{align}
    y_\ell &=  \alpha 
    \left( \sqrt{\frac{\kfactor} {\kfactor+1}} \aa(\pp_\ell)  + \sqrt{\frac{1}{\kfactor+1}} \hhtilde \right)^\top
    \Omegabl \aa(\pp_{b}) + n_\ell,
    \label{eq:obs_model_multipath}
\end{align}
where $\hhtilde \in \mathbb{C}^{M\times1}$ is the \ac{NLoS}\footnote{The \ac{NLoS} is herein defined with respect to the direct path from the \ac{RIS} to the \ac{UE} and accounts for the effect of the uncontrolled multipath components besides this direct path.
}
component of the \ac{RIS}-\ac{UE} channel with $\hhtilde \sim \mtcn(\boldzero, \II)$ and $\kfactor$ is the Rician factor. From~(\ref{eq:obs_model_multipath}), we notice that $\kfactor \rightarrow 0$ results in a Rayleigh channel which translates to losing the \ac{LoS} path between the \ac{RIS} and \ac{UE}. 
Fig.~\ref{fig:multipath} depicts the effect of the multipath profile introduction on the proposed position estimation algorithms, more precisely, Algorithms~\ref{alg:init_pos} and~\ref{alg:ref_pos} as a function of $\kfactor \in [5, 10^3]$, while the \ac{RIS}-\ac{UE} distance is set to $2\unit{m}$. We also plot the \ac{PEB} of the same scenario (in terms of velocity and distance) but without any multipath for the sake of benchmarking. We notice that, 
as $\kfactor$ increases, the \ac{RMSE} levels of both algorithms improve but coincide until around $\kfactor \approx 10^2$ where the refinement routine surpasses the \ac{GS} output and starts operating normally again.
Finally, as $\kfactor \approx 10^3$, a significant drop in the \ac{RMSE} of the refinement algorithm is noticed while that of the \ac{GS} seems to decrease more slowly, but both algorithms follow the trend approaching the \ac{PEB} values. Mitigating a bit the results of the previous sensitivity analysis, and in light of comments from \cite{Iqbal19_FadingModels} regarding the validity of small-scale fading models in the \ac{mmWave} domain, we recall that the combined effects of spatial filtering (thanks to large antenna arrays) and higher power path losses would anyway lead to the reception of relatively sparse and weak secondary multipath components (besides the direct \ac{RIS}-\ac{UE} path) in our case.       
\begin{figure}
    \centering
    \resizebox{1\columnwidth}{!}{
%
%
\definecolor{amber}{rgb}{1.00000,0.75,0.00000}%
\begin{tikzpicture}

\begin{axis}[%
width=4.521in,
height=2.563in,
at={(0.758in,0.484in)},
scale only axis,
xmode=log,
xmin=0.01,
xmax=10000,
xlabel style={font = \large, font=\color{white!15!black}},
xlabel={$\kfactor$},
ymode=log,
ymin=0.0008,
ymax=170,
ylabel style={font = \large, font=\color{white!15!black}},
ylabel={RMSE$\unit{[m]}$},
axis background/.style={fill=white},
xmajorgrids,
xminorgrids,
ymajorgrids,
yminorgrids,
legend style={font = \large, at={(0.51,0.524)}, anchor=south west, legend cell align=left, align=left, draw=white!15!black}
]

\addlegendimage{empty legend}
\addlegendentry{\textbf{With Multipath}}
\addplot [color=red, line width=2.0pt, mark size=5.0pt, mark=asterisk, mark options={solid, red}]
  table[row sep=crcr]{%
0             142.2316\\
0.01          154.5931\\
0.046         146.6900\\
0.215         142.1061\\
1             48.3817\\
4.641         0.5280\\
21.544        0.1911\\
100           0.1007\\
464.158       0.0566\\
2154.434      0.0576\\
10000         0.0494\\
};
\addlegendentry{Grid Search}

\addplot [color=amber, line width=2.0pt, mark size=5pt, mark=square, mark options={solid, amber}]
  table[row sep=crcr]{%
0               139.5882\\
0.01            152.0123\\
0.046           145.1871\\
0.215           139.8541\\
1               47.2036\\
4.641           0.4870\\
21.544          0.1586\\
100             0.0801\\
464.158         0.0432\\
2154.434        0.0434\\
10000           0.0127\\
};
\addlegendentry{CF Position Refinement}

\addlegendimage{empty legend}
\addlegendentry{\textbf{Without Multipath}}

\addplot [color=blue, line width=1.8pt, mark=o, dashed, mark options={solid, blue}, mark size=5.0pt]
  table[row sep=crcr]{%
0               0.00112\\
0.01            0.00112\\
0.046           0.00112\\
0.215           0.00112\\
1               0.00112\\
4.641           0.00112\\
21.544          0.00112\\
100             0.00112\\
464.158         0.00112\\
2154.434        0.00112\\
10000           0.00112\\
};
\addlegendentry{PEB}

\addplot [color=red, line width=2.0pt, dashed, mark size=5.0pt, mark=asterisk, mark options={solid, red}]
  table[row sep=crcr]{%
0             0.0089\\
0.01          0.0089\\
0.046         0.0089\\
0.215         0.0089\\
1             0.0089\\
4.641         0.0089\\
21.544        0.0089\\
100           0.0089\\
464.158       0.0089\\
2154.434      0.0089\\
10000         0.0089\\
};
\addlegendentry{Grid Search}

\addplot [color=amber, line width=2.0pt, dashed, mark size=5pt, mark=square, mark options={solid, amber}]
  table[row sep=crcr]{%
0               0.00128\\
0.01            0.00128\\
0.046           0.00128\\
0.215           0.00128\\
1               0.00128\\
4.641           0.00128\\
21.544          0.00128\\
100             0.00128\\
464.158         0.00128\\
2154.434        0.00128\\
10000           0.00128\\
};
\addlegendentry{CF Position Refinement}

\end{axis}

\end{tikzpicture}%
}
    \caption{\ac{RMSE} of position estimation for the proposed \ac{GS} and \ac{CF} refinement algorithms versus the multipath Rician $\kfactor$-factor. The \ac{RIS}-\ac{UE} distance is set to $\rho = 2\unit{m}$ and the \ac{UE}'s velocity to $v = 1\unit{m/s}$, and $\kfactor \in [5, 10^3]$.}
    \label{fig:multipath}
\end{figure}
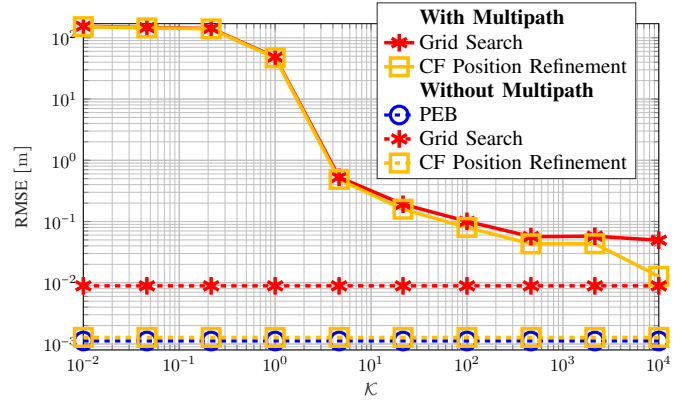

\subsubsection{Sensitivity to SNR}
We also evaluated the performance of our system \ac{w.r.t.}~
    $\text{SNR} = {|\alpha|^2}/({N_0 n_f W})$.
Thus, in Fig.~\ref{fig:SNR} we have plotted the \ac{PEB} performance alongside our devised estimation routines as we change the \ac{SNR}, all while fixing the \ac{RIS}-\ac{UE} distance and \ac{UE} velocity to $5\unit{m}$ and $1\unit{m/s}$ respectively. We first notice that the \ac{PEB} follows an expected trend of linearly decreasing with higher \ac{SNR} values. On the other hand, Algorithm~\ref{alg:init_pos} does not achieve acceptable levels of accuracy at very low \ac{SNR} values, which means that no localization service could be guaranteed at such conditions. However, as the \ac{SNR} hits $-20\unit{dB}$, the performance of the aforementioned algorithm is boosted to the point where it touches the \ac{PEB}, and then continuous at that same level even if the \ac{SNR} is significantly increased. This saturation is due to the finite grid resolution used in this algorithm. 
Finally, it is noticeable that the \ac{CF} position refinement algorithm fail completely at very low \ac{SNR}, but then, also at $-20\unit{dB}$ its  performance comes back and it follows the trend of the \ac{PEB}; and of course, since its output is fed as input to the global refinement algorithm, the latter follows the same trend as well.
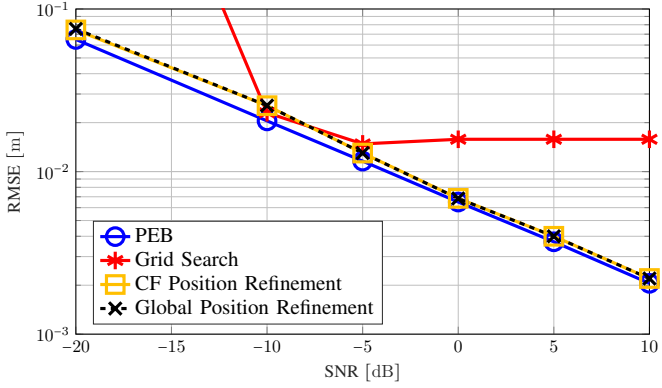
\begin{figure}[!ht]
    \centering
    \resizebox{1\columnwidth}{!}{
%
%
\definecolor{amber}{rgb}{1,0.75,0}
\begin{tikzpicture}

\begin{axis}[%
width=4.521in,
height=2.563in,
at={(0.758in,0.481in)},
scale only axis,
xmin=-20,
xmax=10,
xlabel style={font=\large, font=\color{white!15!black}},
xlabel={SNR$\unit{[dB]}$},
xtick = {-20, -15, -10, -5, 0, 5, 10},
ymode=log,
ymin=0.001,
ymax=0.1,
yminorticks=true,
ylabel style={font=\large, font=\color{white!15!black}},
ylabel={RMSE$\unit{[m]}$},
axis background/.style={fill=white},
xmajorgrids,
ymajorgrids,
yminorgrids,
legend pos=south west,
legend style={font = \large, legend cell align=left, align=left, draw=white!15!black}
]
\addplot [color=blue, line width=1.8pt, mark=o, mark options={solid, blue}, mark size=5.0pt]
  table[row sep=crcr]{%
-20	0.0649780407336089\\
-10	0.0205\\
-5  0.0116\\
0	0.00649780407327552\\
5   0.0037\\
10	0.0020547860661111\\
};
\addlegendentry{PEB}

\addplot [color=red, line width=1.8pt, mark=asterisk, mark options={solid, red}, mark size=5.0pt]
  table[row sep=crcr]{%
-20	14.07\\
-10	0.023\\
-5  0.0148\\
0	0.0158\\
5	0.0158\\
10	0.0158\\
};
\addlegendentry{Grid Search}

\addplot [color=amber, line width=1.8pt, mark=square, mark options={solid, amber}, mark size=5.0pt]
  table[row sep=crcr]{%
-20	0.0743410812747523\\
-10	0.0254287014564301\\
-5  0.013\\
0	0.00684364052896947\\
5   0.004\\
10	0.00220005996357029\\
};
\addlegendentry{CF Position Refinement}

\addplot [color=black, dashed, line width=1.8pt, mark=x, mark options={solid, black}, mark size=5.0pt]
  table[row sep=crcr]{%
-20	0.0754022079036618\\
-10	0.0254577823653587\\
-5  0.013\\
0	0.00683332707449552\\
5   0.004\\
10	0.00219976370331805\\
};
\addlegendentry{Global Position Refinement}

\end{axis}

\end{tikzpicture}%
}
    \caption{\ac{RMSE} of position estimation for the proposed Algorithms~\ref{alg:init_pos} and~\ref{alg:ref_pos} versus the \ac{SNR}. The \ac{RIS}-\ac{UE} distance is set to $\rho = 5\unit{m}$ and the \ac{UE}'s velocity to $v = 1\unit{m/s}$, as a function of the SNR.}
    \label{fig:SNR}
\end{figure}
\subsection{Convergence}
Finally, we empirically illustrate the average convergence rates (over all the Monte Carlo simulation trials) of both Algorithm~\ref{alg:overall} (lines 4 through 7) and Algorithm~\ref{alg:init_pos} (lines 6 through 8) in Fig.~\ref{fig:obj_conv_a} and~\ref{fig:obj_conv_b}, respectively, where the \ac{RIS}-\ac{UE} distance is set to $2\unit{m}$ and the \ac{UE}'s velocity is set to $1\unit{m/s}$. 
Convergence is considered to be accomplished if the evolution of the respective objective functions~(\ref{eq:1D_distance}) and~(\ref{eq:ml_pos_vel_gain}) is smaller than an arbitrarily chosen threshold $e\approx10^{-15}$.
Fig.~\ref{fig:obj_conv_a} shows that~\eqref{eq:1D_distance} converges in just $3$ iterations whereas Fig.~\ref{fig:obj_conv_a} shows that~\eqref{eq:ml_pos_vel_gain} takes around $20$ iterations to finally converge, which is still acceptable from an algorithmic point-of-view. This practically shows how quick the two algorithmic steps can be in a typical application case.
Moreover, Algorithms~\eqref{alg:ref_pos} and~\ref{alg:ref_vel} converge once the objective functions in~\eqref{eq:ml_pos_gain_appr} and \eqref{eq:ml_vel_gain_appr} converge to the prespecified threshold, meaning that the error between the received observation and the one reconstructed via the updated position and velocity residuals (respectively) tends to $0$.


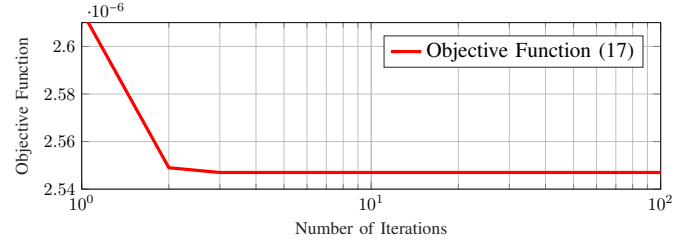
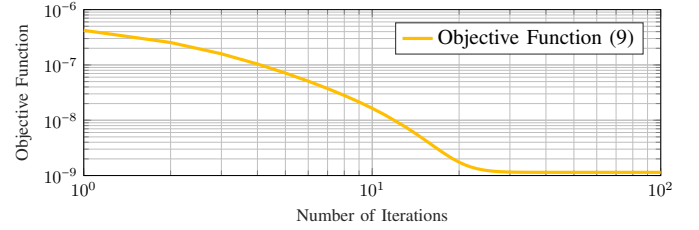
\begin{figure}
     \centering
     \label{}
     \begin{subfigure}[b]{1\columnwidth}
         \centering
         \resizebox{1\columnwidth}{!}{
%
\definecolor{amber}{rgb}{1.00000,0.75,0.00000}%

\begin{tikzpicture}
\begin{axis}[%
width=4.521in,
height=1.3in,
at={(0.758in,0.481in)},
scale only axis,
xmin=1,
xmax=100,
xlabel style={font = \large, font=\color{white!15!black}},
xlabel={Number of Iterations},
xmode=log,
ymin=2.54e-06,
ymax=2.61e-06,
ylabel style={font = \large, font=\color{white!15!black}},
ylabel={Objective Function},
axis background/.style={fill=white},
xmajorgrids,
ymajorgrids,
yminorgrids,
xminorgrids,
legend style={font = \large, at={(0.97,0.73)}, anchor=south east, legend cell align=left, align=left, draw=white!15!black}
]
\addplot [color=red, line width=1.8pt]
  table[row sep=crcr]{%
1	2.615e-06\\
2	2.549e-06\\
3	2.547e-06\\
4	2.547e-06\\
5	2.547e-06\\
6	2.547e-06\\
7	2.547e-06\\
8	2.547e-06\\
9	2.547e-06\\
10	2.547e-06\\
11	2.547e-06\\
13	2.547e-06\\
14	2.547e-06\\
15	2.547e-06\\
16	2.547e-06\\
17	2.547e-06\\
18	2.547e-06\\
19	2.547e-06\\
20	2.547e-06\\
21	2.547e-06\\
22	2.547e-06\\
23	2.547e-06\\
24	2.547e-06\\
25	2.547e-06\\
26	2.547e-06\\
27	2.547e-06\\
28	2.547e-06\\
30	2.547e-06\\
32	2.547e-06\\
36	2.547e-06\\
46	2.547e-06\\
100 2.547e-06\\
};
\addlegendentry{Objective Function~\eqref{eq:1D_distance}}
\end{axis}

\end{tikzpicture}%
}
         \caption{Convergence of  Algorithm~\ref{alg:init_pos} via objective function~\eqref{eq:1D_distance} }
        \label{fig:obj_conv_a}
    \end{subfigure}
     \hfill
     \begin{subfigure}[b]{1\columnwidth}
         \centering
         \resizebox{1\columnwidth}{!}{
%
\definecolor{amber}{rgb}{1.00000,0.75,0.00000}%

\begin{tikzpicture}
\begin{axis}[%
width=4.521in,
height=1.3in,
at={(0.758in,0.481in)},
scale only axis,
xmin=1,
xmax=100,
xlabel style={font = \large, font=\color{white!15!black}},
xlabel={Number of Iterations},
ymode=log,
xmode=log,
ymin=1e-09,
ymax=1e-06,
yminorticks=true,
ylabel style={font = \large, font=\color{white!15!black}},
ylabel={Objective Function},
axis background/.style={fill=white},
xmajorgrids,
ymajorgrids,
yminorgrids,
xminorgrids,
legend style={font = \large, at={(0.97,0.73)}, anchor=south east, legend cell align=left, align=left, draw=white!15!black}
]
\addplot [color=amber, line width=1.8pt]
  table[row sep=crcr]{%
1	4.21784065056705e-07\\
2	2.54264378754792e-07\\
3	1.58855826797792e-07\\
4	1.03895808273926e-07\\
5	7.1268711115622e-08\\
6	5.08805293583305e-08\\
7	3.74108953599526e-08\\
8	2.80829153423492e-08\\
9	2.13819239870115e-08\\
10	1.6431273622503e-08\\
11	1.26970582366914e-08\\
13	7.64131822522012e-09\\
14	5.94676430181896e-09\\
15	4.64932999425744e-09\\
16	3.667390053599e-09\\
17	2.93579584028949e-09\\
18	2.40046264062052e-09\\
19	2.01587032684449e-09\\
20	1.74417722281482e-09\\
21	1.55486777213491e-09\\
22	1.42427074791603e-09\\
23	1.33474551854241e-09\\
24	1.27358169941648e-09\\
25	1.23185534410129e-09\\
26	1.20340274613538e-09\\
27	1.18400193707681e-09\\
28	1.17077402718736e-09\\
30	1.15560717637918e-09\\
32	1.14856190981162e-09\\
36	1.14377546926759e-09\\
46	1.14249329841477e-09\\
100	1.14246525608524e-09\\
};
\addlegendentry{Objective Function~\eqref{eq:ml_pos_vel_gain}}
\end{axis}

\end{tikzpicture}%
}
         \caption{Convergence of Algorithm~\ref{alg:overall} via objective function~\eqref{eq:ml_pos_vel_gain} }
        \label{fig:obj_conv_b}
     \end{subfigure}
     \caption{Convergence rate of Algorithms~\ref{alg:overall} and~\ref{alg:init_pos}, via objective functions~\eqref{eq:ml_pos_vel_gain} and~\eqref{eq:1D_distance} respectively, as a function of the number of iterations at $\rho = 2\unit{m}$ \ac{RIS}-\ac{UE} distance and $v= 1\unit{m/s}$ \ac{UE}'s velocity.}
\end{figure}
%

\section{Conclusion}
\label{sec:Conclusion}
In this work, we have proposed a single-\ac{BS} \ac{DL} \ac{SISO} positioning scenario, by considering the 6D state estimation of a mobile \ac{UE}, while relying on one reflective \ac{RIS} in the presence of \ac{LoS} blockage. 
We have accounted for small-scale velocity effects by modeling the \ac{RIS} response in geometric \ac{NF} (i.e., taking into account extra phase shifts at the \ac{RIS} elements); and we have derived both theoretical position and velocity error bounds to be used as performance benchmarks. Then, we have designed a 6D estimation algorithm consisting of three subroutines. First, we assume the \ac{UE} to be static and we coarsely estimate the position parameters via grid search which are then passed to an iterative subroutine that refines the results while assuming the knowledge of the \ac{UE}'s velocity. Then, we estimate the velocity parameters with the perfect knowledge assumption of the \ac{UE}'s position. 
We have introduced algorithmic time complexity analysis and we have simulated our algorithms against different system parameters, including the \ac{RIS}-\ac{UE} distance and \ac{UE}'s velocity. The obtained results show that our final algorithmic results attain the theoretical performance bounds at close \ac{RIS}-\ac{UE} distances. For farther distances however, as our \ac{CF} refinement routines are shown to experience much poorer performance, we have resolved this issue by introducing a global refinement process (as part of our algorithm), which restores a fine performance level. Furthermore, we have tested the resilience of our algorithm against both low \ac{SNR} values and multipath interference.

This work is to be extended in multiple dimensions, including more realistic propagation models (e.g., channel nonstationarity, time-varying directions from the RIS) and 
\ac{UE} state tracking through Bayesian filtering. Moreover, previous research works could be leveraged to include location-based \ac{RIS} phase optimization routines, as well as hardware constraints of real \ac{RIS} prototypes. 
Combining all these blocks within one unified system approach (i.e., snapshot positioning with localization-optimal \ac{RIS} control, followed by a tracking filter) is expected to result in better snapshot estimation, hence better filter observations and ultimately, even better corrections beyond mobility-based predictions to anticipate on the best RIS configurations for next snapshot estimation steps (up to any arbitrary time horizon).
\\
\subsubsection*{Acknowledgment}
The authors would like to express their sincere gratitude to Dr. Thomas Zemen from the Austrian Institute of Technology (AIT), Austria, for his valuable feedback and insightful suggestions, which helped improve the clarity and accuracy of several expressions in this work.


\appendices
\section{Derivation of $\flm(\pp,\vv)$ in \eqref{eq:flm}}
\label{app:flm}
In this part, we show the derivation of the approximation \eqref{eq:flm} from the original expression \eqref{eq:flm_undeveloped}. Introducing $\deltam = {\pp_m - \pp}$ and $\deltaell = \vv \ell \Ts$, the first term in \eqref{eq:flm_undeveloped} can be re-cast as
\begin{align}
    \norm{\pp + \vv \ell T_s - \pp_m} &= \norm{\deltam - \deltaell }
     = \norm{\deltam}
    \frac{\norm{\deltam - \deltaell } }{\norm{\deltam}} ~,\\
    & = \norm{\deltam}
    \sqrt{\frac{{(\deltam - \deltaell)^\top (\deltam - \deltaell)}}{\norm{\deltam}^2} }~,\\
    & = \norm{\deltam}
    \sqrt{1
    - \frac{2 \bm{\delta}_m^\top \deltaell}{\norm{\bm{\delta}_m}^2}
    + \frac{\norm{\deltaell}^2 }{\norm{\bm{\delta}_m}^2}
    }~. \label{eq:approx_3}
\end{align}

We now assume that the total displacement of the \ac{UE} during the entire observation interval is negligible compared to the distance between the \ac{RIS} and the \ac{UE}, i.e., 
\begin{align}
\frac{\norm{\bm{\Delta}_{\ell}}^2 }{\norm{\bm{\delta}_m}^2} \ll 1, \forall m, \ell, \label{eq:d_limit}
\end{align}
or, equivalently\footnote{Note that the scenario we consider in this work, including the mobility assumptions, comply with this condition as explained in~\ref{sec:scen_def_and_perf_met}.}
\begin{align}
    \norm{\vv} &\ll \frac{\norm{\bm{\delta}_m}}{L\Ts}, \forall m. \label{eq:v_limit}
\end{align}



Based on~\eqref{eq:approx_3} and~\eqref{eq:d_limit}, we consider the function 
%
$f(x) = \sqrt{1 - 2x + e}$, where $x \triangleq \frac{\bm{\delta}_m^\top\bm{\Delta}_{\ell}}{\norm{\bm{\delta}_m}^2} $ and $e \triangleq \frac{\norm{\bm{\Delta}_{\ell}}^2 }{\norm{\bm{\delta}_m}^2}$ is  an arbitrarily small quantity, which yields the approximation $f(x) \approx \sqrt{1 - 2x}$. In addition, the condition in \eqref{eq:v_limit} also implies that (considering $\norm{\deltam - \deltaell} \gg \norm{\deltaell}$)
\begin{align}
\frac{2 \bm{\delta}_m^\top\bm{\Delta}_{\ell}}{\norm{\bm{\delta}_m}^2} \ll 1 ~. \label{eq:d_limit2}    
\end{align}
Following~\eqref{eq:d_limit2}, we apply the first order Taylor's expansion to the approximated $f(x)$ around $x=0$, which yields $f(x) \approx 1 - x$.
%
Hence, \eqref{eq:approx_3} can be approximated as 
\begin{align} \nonumber
    \norm{\pp + \vv \ell T_s - \pp_m}
    & \approx \norm{\bm{\delta}_m} - \frac{\bm{\delta}_m^\top\bm{\Delta}_{\ell}}{\norm{\bm{\delta}_m}} ~,
    \\ \nonumber
    & = \norm{\pp_m - \pp} + \frac{(\pp - \pp_m)^\top}{\norm{\pp_m - \pp}}
    \vv\ell\Ts ~,
    \\ \label{eq_first_approxx}
    &= d_m + \uu_{m}^\top(\pp) \vv \ell \Ts ~.
\end{align}
Similarly, the second term in \eqref{eq:flm_undeveloped} can be approximated as
\begin{align} \label{eq_second_approxx}
    \norm{\pp - \pp_r}
    & \approx 
     d_r  ~.
\end{align}
Combining \eqref{eq_first_approxx} and \eqref{eq_second_approxx} yields \eqref{eq:flm}.

\section{\ac{FF} Approximation}
\label{app:FF_approximation}
The NF RIS response form in~\eqref{eq:nf_steer} can be in fact approximated to its equivalent FF form as following.\\
From the geometry of Fig.~\ref{fig:model} we have:
\begin{align}
d_{m} = \sqrt{\dr^{2} + \qm^{2} - 2\dr \qm\kappa
} ~,
\end{align}
where $\qm = \norm{\pp_m - \pp_r}$ and $\kappa \triangleq \sin{(\phi)}\cos{(\theta - \psi_{m})} $ is the constant holding the angular (azimuth and elevation) terms.\\
Using first-order Taylor's expansion, the function $f(x) = \sqrt{1 + x^{2} - 2x\kappa}$ can be approximated around $x = 0$ into $f(x) \approx 1 - x\kappa$. Then given that the RIS is large enough and the \ac{UE} is far from the surface, we can safely assume that $q_m \ll d_r$,  which allows us to approximate $d_{m}$ as
\begin{align}
 d_{m} & =  d_{r} \sqrt{1 + \frac{q_{m}^{2}}{d_{r}^{2}} - 2\frac{q_m}{d_r}\kappa}\\
 d_{m}  & \approx \dr(1 - \frac{\qm}{\dr}\kappa) \label{eq:dm_approx}\\
        &= \dr - \qm\kappa,
\end{align}
then
\begin{align}
    \norm{\pp_{m} - \pp} -
    \norm{\ppr - \pp} = \dm - \dr = -\qm \sin{(\phi)}\cos{(\theta - \psi_{m})}.
\end{align}
As mentioned above, the condition for the approximation in~\eqref{eq:dm_approx} to hold is $q_m \ll d_r$. This means that the distance from the \ac{RIS} reference element to the \ac{UE} should be at least around 10 times bigger than the distance between the former to the $m$-th \ac{RIS} element. As an example, consider that the \ac{UE} is at the shortest considered distance to the \ac{RIS} in this study, thus $d_r = 1\unit{m}$. Moreover, the \ac{RIS} used in this study is a square surface of size $32\times32$ elements separated by a distance of $\frac{\lambda}{2}$, which means that the biggest $q_m$ is $\approx 0.1174\unit{m}$. Note that, this \ac{NF} to \ac{FF} approximation depends on both $q_m$ and $d_r$, so if the \ac{RIS} is electronically bigger, this yields a bigger $q_m$ and hence the \ac{UE} must get further from the surface for this approximation to hold.
Following \cite{abu2020near}, we then approximate the steering vector in \ac{FF} as
\begin{align} \label{eq:ff}
    [\aa(\theta, \phi)]_{m} \approx 
    \exp{
    (- j (\pp_{m} - \ppr)^{\top}\kk(\phi, \theta))
    } ~,
\end{align}
where
$\kk(\phi,\theta) = -\frac{2\pi}{\lambda}[\sin{\phi}\cos{\theta}, \hspace{1mm}
\sin{\phi}\sin{\theta}, \hspace{1mm}
\cos{\phi}]^\top$.

\section{Gradient of $\flm(\pp,\vv)$}
\label{app:gradient}

In the linearized RIS response model from \eqref{eq:nf_steer_appr}, we have that
\begin{align}\label{eq:grad_flm}
\gradF(\pp, \vv) &= \frac{\pp_m-\pp}{\dm}-\frac{\ppris-\pp}{\dr} \nonumber \\
&+  \left( \partder{\uu_m(\pp)}{\pp} \right)^\top  \vv\ell\Ts \nonumber \\
&= \uu_m(\pp) - \uuris(\pp) + \gradGamT \vv\ell\Ts \in \realset{3}{1}~,\\
\gradGam 
    &\triangleq \left(
    \frac
     {\norm{\pp_m-\pp}^2 \boldone_{3\times3} - (\pp_m-\pp)(\pp_m-\pp)^\top }
    {\norm{\pp_m-\pp}^3} \right) \\
    & \in \realset{3}{3}. \notag
\end{align}

\section{Derivation of $\pd$}
\label{app:pd}

The problem \eqref{eq:ml_pos_gain_appr} for fixed $\alpha = \alphahat$ can be reduced to
\begin{align}\label{eq:ml_pos_vel_gain_appr_pd}
	\pdhat = \arg \min_{\pd \in \realset{3}{1}}  \norm{\yy - \alphahat \,  (\etab + j \bxib^\top \pd )  }^2 ~.
\end{align}
Opening up the terms in \eqref{eq:ml_pos_vel_gain_appr_pd} ~, we have
\begin{align} \nonumber
    \llk(\pd) &\triangleq \norm{\yy - \alphahat \,  (\etab + j \bxib^\top \pd )  }^2 ~,
    \\ \nonumber
    &= \norm{\yy}^2 - 2 \realp{\yy^\her \alphahat \,  (\etab + j \bxib^\top \pd ) } + \abs{\alphahat}^2 \norm{\etab + j \bxib^\top \pd}^2 ~,
    \\ \nonumber
    &= \norm{\yy}^2 - 2 \realp{ \alphahat \, \yy^\her  \etab } - 2 \realp{j \alphahat \, \yy^\her  \bxib^\top \pd  } + \abs{\alphahat}^2 \norm{\etab}^2 \\ \nonumber
    & + 2 \abs{\alphahat}^2 \realp{j \etab^\her  \bxib^\top \pd} + \abs{\alphahat}^2 \norm{j \bxib^\top \pd}^2 ~,
    \\ \label{eq:lpd}
    &= \abs{\alphahat}^2 \pd^\top \bxib^\conj \bxib^\top \pd + \big( 2 \abs{\alphahat}^2 \realp{j \etab^\her  \bxib^\top} \nonumber \\ 
    & - 2 \realp{j \alphahat \, \yy^\her  \bxib^\top  } \big) \pd 
 +\norm{\yy}^2 \nonumber \\ 
    & - 2 \realp{ \alphahat \, \yy^\her  \etab } +\abs{\alphahat}^2 \norm{\etab}^2,
\end{align}
where we have used the fact that $\pd$ is a real vector. According to Lemma~S1 in the supplementary material of \cite{PN_Radar_2023}, $\pd^\top \bxib^\conj \bxib^\top \pd = \pd^\top \realp{\bxib^\conj \bxib^\top} \pd$ since $\pd \in \realset{3}{1}$ and $\bxib^\conj \bxib^\top$ is Hermitian. Then, ignoring the constant terms in \eqref{eq:lpd}, the problem \eqref{eq:ml_pos_vel_gain_appr_pd} can be re-written as
\begin{align}
	\pdhat &= \arg \min_{\pd \in \realset{3}{1}}  \left\{ \abs{\alphahat}^2 \pd^\top \realp{\bxib^\conj \bxib^\top} \pd \right. \\
 & \left. +  \big( 2 \abs{\alphahat}^2 \realp{j \etab^\her  \bxib^\top} - 2 \realp{j \alphahat \, \yy^\her  \bxib^\top  } \big) \pd \right\} ~. \label{eq:ml_pos_vel_gain_appr_pd2}
\end{align}
Since \eqref{eq:ml_pos_vel_gain_appr_pd2} is an unconstrained quadratic optimization problem, the solution can readily be obtained in closed-form as
\begin{align} \label{eq:pdhat_2}
    \pdhat = \frac{1}{\abs{\alphahat}^2} \big( \realp{\bxib^\conj \bxib^\top} \big)^{-1} \imp{ \bxib \big(    \abs{\alphahat}^2 \etab^\conj   -  \alphahat \, \yy^\conj  \big)    } ~.
\end{align}
Notice 
that the $3 \times 3$ matrix
\begin{align}
    \realp{\bxib^\conj \bxib^\top} = \sum_{\ell=1}^{L} \realp{ \bxi_{\ell}^\conj \bxi_{\ell}^\top }
\end{align}
is full-rank for $L \geq 3$ under mild conditions (e.g., if RIS phase profiles are random). Hence, if the number of transmissions is at least $3$, then the closed-form UE position can be computed using~\eqref{eq:pdhat}.

\textit{Update $\alpha$ for fixed $\pd$:} The solution of $\alpha$
for a fixed $\pd = \pdhat$ is given by
\begin{align} \label{eq:alphahat_p_2}
    \alphahat = \frac{ (\etab + j \bxib^\top \pdhat )^\her \yy }{ \norm{\etab + j \bxib^\top \pdhat}^2 } ~.
\end{align}

\section{PEB \& FIM Derivations}
\label{app:PEB}
We herein derive the full \ac{FIM} and \ac{PEB} with unknown position and velocity. The vector of unknown parameters to be estimated is defined as
\begin{align}
    \zeta &= [\pp^\top, \vv^\top, \alpha_r, \alpha_i]^\top\\
    & = [p_x, p_y, p_z, v_x, v_y, v_z, \alpha_r, \alpha_i]^\top \in \mathbb{R}^{8\times1},
\end{align}
which yields an $8\times8$ \ac{FIM} expressed as
\begin{align}
    \text{\ac{FIM}} = \frac{2}{P_n}\realp{\left(\frac{\partial\mu}{\partial\zeta}\right)^H \frac{\partial\mu}{\partial\zeta}} \in \mathbb{R}^{8\times8},
\end{align}
where $\mu_\ell = \alpha\omega_\ell^\top \aa(\pp_\ell)$ and
\begin{align}
    \frac{\partial\mu}{\partial\zeta} = 
    \begin{bmatrix}
     \frac{\alpha\omega_\ell^\top \partial\aa(\pp_\ell)}{\partial\pp}\vspace{2mm}\\
     \frac{\alpha\omega_\ell^\top
    \partial\aa(\pp_\ell)}{\partial\vv}\vspace{2mm}\\
     \alpha_i\omega_\ell^\top \aa(\pp_\ell)\vspace{2mm}\\
     \alpha_r\omega_\ell^\top \aa(\pp_\ell)
    \end{bmatrix}^\top.
\end{align}
The the \ac{PEB} and \ac{VEB} expressions can be derived as
\begin{align}
    \ac{PEB} & = \sqrt{\text{tr}\left( [\ac{FIM}^{-1}]_{1:3,1:3} \right)}\\
    \ac{VEB} & = \sqrt{\text{tr}\left( [\ac{FIM}^{-1}]_{4:6,4:6} \right)}.
\end{align}

\balance 
\bibliographystyle{IEEEtran}
\bibliography{references}
\end{document}